\documentclass[aps,pre,preprint,onecolumn,citeautoscript,superscriptaddress,eqsecnum]{revtex4-1}  
\usepackage[utf8]{inputenc}
\usepackage{epsfig}
\usepackage{graphicx}
\usepackage{dcolumn}
\usepackage{caption}
\usepackage{subcaption}
\usepackage{hyperref}
\usepackage{ulem}
\usepackage{bm}
\usepackage{color}
\usepackage{braket}
\usepackage{latexsym}
\usepackage{amsmath}
\usepackage{amssymb}
\hypersetup{bookmarks,
            colorlinks,
           citecolor=red,
            filecolor=blue,
           urlcolor=blue,
           pdfpagemode=None}
\usepackage{latexsym}
\usepackage{todonotes}

\def \beq{\begin{eqnarray}}
\def \eeq{\end{eqnarray}}
\def \r{{\mathbf{r}}}

\def \v{{\mathbf{v}}}
\def \x{{\mathbf{x}}}
\def \y{{\mathbf{y}}}
\def \j{{\mathbf{j}}}

\def \Q{{\mathbf{Q}}}
\def \K{{\mathbf{K}}}
\def \q{{\mathbf{q}}}
\def \k{{\mathbf{k}}}
\def \vd{{\mathbf{v}_{\Delta}}}
\def \vf{{\mathbf{v}_F}}

\def \xh{\hat{x}}
\def \yh{\hat{y}}
\def \ua{\uparrow}
\def \da{\downarrow}
\def \nn{\nonumber \\}
\newcommand{\bs}{\boldsymbol}
\newcommand{\ie}{\textit{i.\,e.}}
\newcommand{\etal}{\textit{et~al.}}

\begin{document}
\title{Thermal and electrical transport in metals and superconductors\\ 
across antiferromagnetic and topological quantum transitions}

\author{Shubhayu Chatterjee}
\affiliation{Department of Physics, Harvard University, Cambridge Massachusetts
02138, USA.}

\author{Subir Sachdev}
\affiliation{Department of Physics, Harvard University, Cambridge Massachusetts
02138, USA.}
\affiliation{Perimeter Institute for Theoretical Physics, Waterloo, Ontario, 
Canada N2L 2Y5}

\author{Andreas Eberlein}
\affiliation{Department of Physics, Harvard University, Cambridge Massachusetts
02138, USA.}

\date{\today \\
\vspace{1.6in}}

\begin{abstract}
We study thermal and electrical transport in metals and superconductors near a 
quantum phase transition where antiferromagnetic order 
disappears. The same theory can also be applied to quantum phase transitions
involving the loss of certain classes of intrinsic topological order.
For a clean superconductor, we recover and extend 
well-known universal results. The heat conductivity for commensurate and 
incommensurate antiferromagnetism coexisting with superconductivity shows a 
markedly different doping dependence near the quantum critical point, thus 
allowing us to distinguish between these states. In the dirty limit, the results for the 
conductivities are qualitatively similar for the metal and the superconductor. 
In this regime, the geometric properties of the Fermi surface allow for a very good
phenomenological understanding of the numerical results on the conductivities.
In the simplest model, we find that the conductivities do not track 
the doping evolution of the Hall coefficient, in contrast to recent 
experimental findings. We propose a doping dependent scattering rate, possibly due 
to quenched short-range charge fluctuations below optimal doping,
to consistently describe both the Hall data and the longitudinal conductivities.
\end{abstract}

\maketitle

\tableofcontents

\section{Introduction}
Recent experimental results on the Hall coefficient in hole doped 
cuprates~\cite{Badoux2016} suggest the existence of a quantum critical point 
(QCP) near optimal doping, at which the charge-carrier density changes by one 
hole per $\mathrm{Cu}$ atom. Results consistent with such a 
scenario were also found in the 
electrical~\cite{Laliberte2016-arXiv,Collignon2016} and 
thermal~\cite{Michon2017} conductivities of various cuprate materials. 
Such a change of the carrier density with decreasing hole doping could be 
caused by various QCPs. In one scenario, the appearance of long-range 
commensurate antiferromagnetic (AF)~\cite{Storey2016}, incommensurate 
antiferromagnetic~\cite{Eberlein2016} or charge-density wave (CDW) 
order~\cite{Sharma2017} leads to a reconstruction of the Fermi surface. In an 
alternative scenario, the QCP is associated with the appearance of a pseudogap 
metal with topological 
order~\cite{Yang2006,Kaul2007,YQi2010,SSDC16,Eberlein2016,Chatterjee2016}. At finite 
temperature, a suppression of the Hall number could also be obtained as a result 
of strongly anisotropic scattering by dynamical CDW 
fluctuations~\cite{Caprara2016-arXiv}.

Ideally, one would like to resolve the Fermi surface on both sides of the QCP 
with spectroscopic probes like ARPES, or using quantum oscillation 
measurements. In the 
underdoped 
regime the former resolves arcs, and it is not clear whether these are closed 
into 
Fermi pockets. Quantum oscillation experiments are difficult in the 
underdoped cuprates due to restrictions on the sample quality or accessible 
temperatures and magnetic fields. In the underdoped regime, quantum 
oscillations have only been observed near a doping of 
$1/8$ hole per $\mathrm{Cu}$ site, where the ground state shows charge-density 
wave order in high magnetic 
fields~\cite{TWu2011,TWu2013,LeBoeuf2013,Gerber2015,Doiron-Leyraud2007,
LeBoeuf2007,Kemper2016,Chan2016}.

Given the lack of direct evidence, it is desirable to further explore the
consequences of different proposals for the QCP at optimal doping and make 
predictions for feasible measurements. Changes in the Hall coefficient are very 
similar in all proposals involving a reconstruction of the large Fermi surface 
into small Fermi pockets with decreasing doping. In this paper, we therefore 
present a detailed discussion of transport properties near a QCP where static 
or fluctuating antiferromagnetic order disappear. The former could be due to 
commensurate or incommensurate spin-density wave order. The latter is 
associated with Fermi liquids with a certain class of topological order \cite{SSDC16}: as we will describe
in Sec.~\ref{sec:topo}, these have transport properties very 
similar to those of conventional Fermi liquids with magnetic order at low 
temperature.

Transport properties of $d$-wave superconductors have mostly been studied in the 
clean limit. In this case, thermal transport is universal in the sense that the 
thermal conductivity depends only on the number of nodes, the Fermi velocity 
and the gap velocity~\cite{DurstLee_PRB2000}. Some cuprate materials are, 
however, not in the clean limit around optimal doping. It is therefore 
interesting to complement studies of the clean limit by the dirty limit, and in the presence of additional symmetry-breaking order parameters. 

This paper is organized as follows. In Sec.~\ref{formalism}, we introduce our most general Hamiltonian, and derive expressions for the Green's function and the thermal current and conductivity from linear response theory. Then, in Sec.~\ref{AFM}, we discuss analytic and numerical results for the conductivity across the antiferromagnetic 
QCP in the metallic limit. We extend our results to include additional superconductivity in Sec.~\ref{AFMandDSC}, discuss the clean and dirty limits, and also make connections with the universal Durst-Lee formula \cite{DurstLee_PRB2000}. In Sec.~\ref{sec:DopingDependentScattering}, we extend our analysis for the dirty superconductor to include a  phenomenological doping-dependent scattering rate, and find good qualitative agreement of the longitudinal conductivities and Hall angle with recent transport experiments \cite{Collignon2016,Michon2017}. We present an alternate model of the pseudogap phase as a topological metal in Sec.~\ref{sec:topo}, and argue that a Higgs transition across a topological QCP results in identical charge and energy transport. We end with a discussion of the effect of additional excitations and fluctuations beyond our mean-field picture of the transition on the transport properties in Sec.~\ref{sec:dis}, and a summary of our main results in Sec.~\ref{sec:conc}.

\section{Model and formalism}
\label{formalism}
\subsection{Hamiltonian}
\label{sec:Hamiltonian}
We consider a mean-field Hamiltonian describing the coexistence and competition 
of superconductivity and spiral 
antiferromagnetism~\cite{Sushkov2004,Yamase2016} in the presence of disorder,
\beq
H &=& H_{0} + H_{SC} + H_{AF}  + H_{dis} \nn
H_0 &=& \sum_{\k} \xi_{\k} c^{\dagger}_{\k \sigma} c_{\k \alpha} \nn
H_{SC} &=&  \sum_{\k} \Delta_{\k} \left( c^{\dagger}_{\k \ua} c^{\dagger}_{-\k 
\da} + c_{-\k \da} c_{\k \ua} \right) \nn
H_{AF} & = &  - \sum_{i} \mathbf{m}_i \cdot \mathbf{S}_i =  - A \sum_{\k} 
\left( 
c^{\dagger}_{\k \ua} c_{\k + \Q \da}  + c^{\dagger}_{\k + \Q \da} c_{\k \ua} 
\right),
\label{Htot}
\eeq
where $\xi_\k = -2 t (\cos k_x + \cos k_y) - 4 t' \cos k_x \cos 
k_y - \mu$ is the fermionic dispersion, $\Delta_{\k} = 
\Delta_d(\text{cos}k_x - \text{cos}k_y)$ is the superconducting pairing gap 
with $d$-wave symmetry and $\mathbf{m}_i = 2 A \left[ \xh \cos(\Q\cdot \r_i)+  
\yh \sin(\Q\cdot \r_i) \right]$ is the in-plane local magnetization that 
corresponds to N\'eel order if the ordering wave vector $\Q = (\pi, 
\pi)$ is commensurate and spiral order for incommensurate $\Q = (\pi 
- 2 \pi \eta, \pi)$. In the following we set $t = 1$ and use it as the unit of 
energy. $H_{dis}$ describes impurity scattering of the electrons, the 
effects of which will be taken into account by a finite scattering rate $\Gamma 
= (2\tau)^{-1}$, where $\tau$ is the quasi particle lifetime of the low-energy 
electrons. 

We evaluate the thermal conductivity in various regimes, including metals with 
commensurate or incommensurate fluctuating or long-range antiferromagnetic 
order, or superconductors in the presence of the latter two orders. We 
distinguish between the clean limit in which $\Gamma \ll \Delta_d$, and the 
dirty limit $\Gamma \gg \Delta_d$. In the clean limit, transport is dominated by 
contributions from the nodes, while in the dirty limit the nodal structure is 
washed out and the entire Fermi surface contributes to transport. Throughout our 
computation, we assume that both $\Gamma$ and $\Delta_d$ are much smaller than 
the Fermi energy $E_F$. Except close to the QCP, they are also significantly 
smaller than the antiferromagnetic gap.

For simplicity, we first choose $\Gamma$ to be independent of doping. We find that while the conductivity drops below the transition, in the dirty limit the drop relative to the phase with no magnetic order is quite small. We note that a self-consistent computation of $\Gamma$ would involve the density of states for the appropriate Fermi surface (the reconstructed one below the critical doping), and a spin-dependent scattering matrix element as the quasiparticles have spin-momentum locking after Fermi surface reconstruction (in case of the long-range magnetic order). The density of states at the Fermi surface decreases gradually across the transition. Further, the scattering matrix element averaged over the Fermi surface decreases in the ordered phase because the smaller overlap between the spin-wavefunctions of initial and final scattering state of the quasiparticle. Hence, these effects cannot further decrease the conductivities on the ordered side, in contradiction with experiments. However, the small Fermi pockets are susceptible to charge density waves, and quenched disorder in form of charge fluctuations can lead to an increased scattering rate. Therefore, we modify our results to have a doping-dependent $\Gamma$ that increases below the critical point, and find that this can consistently explain both the Hall data and the longitudinal conductivities in the dirty limit.

\subsection{Thermal current operator}
The details of the computation of the thermal current operator and thermal 
conductivity depend upon the state and limit under consideration. We first 
derive the most general thermal current operator for the models considered, and 
outline our approach to evaluating the thermal conductivity via the Kubo 
formula.

We generalize the derivation of the heat current operator for a $d$-wave 
superconductor presented in \cite{DurstLee_PRB2000} to include additional 
magnetic order. In presence of co-existing charge density wave order and 
superconductivity, the heat current operator can be derived from the spin 
current operator, as quasiparticles have conserved 
$s_z$~\cite{DurstSachdev_PRB2009}. However, the Hamiltonian in 
Eq.~\eqref{Htot} does not possess the $U(1)$ symmetry corresponding to the 
conservation of $S_z$, so we need to derive the thermal current operator from 
scratch. We do so for a general value of $\Q$, so that our results also apply 
for the incommensurate case. 

We work in the continuum limit in position space. At the end of the 
computation, we can replace the Fermi velocity by that of the lattice model, 
which is equivalent to neglecting interband contributions in the presence of 
magnetic order. This approximation has been used before for the computation of 
the electrical and Hall conductivities of spiral antiferromagnetic 
states~\cite{Voruganti1992,Eberlein2016}. Moreover, we assume that the pairing 
amplitude $\Delta_d$ is real. The Hamiltonian for the clean system is then given 
by
\beq
H &=& \frac{1}{2 m} \int d\x \, \nabla c^{\dagger}_{\alpha}(\x). \nabla c_{\alpha}(\x) + \int d\x \, d\y \; \Delta(\x - \y) \left[ c^{\dagger}_{\ua}(\x) c^{\dagger}_{\da}(\y) + c_{\da}(\y) c_{\ua}(\x) \right] \nn
 & & - A \int d\x \, \left[ e^{ -i \Q\cdot\x} c^{\dagger}_{\ua}(\x) c_{\da}(\x) + e^{i \Q\cdot \x} c^{\dagger}_{\da}(\x) c_{\ua}(\x) \right] \equiv  \int d\x \, h(\x), 
\eeq 
where $h(\x)$ is the local Hamiltonian density. The latter can be identified 
with the heat density if we measure energies with respect to the chemical 
potential. Therefore, the thermal current operator $\j^{Q}(\x)$ can be defined 
by the continuity equation:
\beq
\dot{h}(\x) + \nabla \cdot \j^{Q}(\x) = 0
\eeq

The time-derivative of the Hamiltonian density is given by
\beq
\dot{h}(\x) & = & \frac{1}{2m} \left[  \nabla \dot{c}^{\dagger}_{\alpha}(\x). \nabla c_{\alpha}(\x) +  \nabla c^{\dagger}_{\alpha}(\x). \nabla \dot{c}_{\alpha}(\x)  \right]  \nn
&  & + \int d\y \, \Delta(\x - \y) \left[  \dot{c}^{\dagger}_{\ua}(\x) c^{\dagger}_{\da}(\y) +  c^{\dagger}_{\ua}(\x) \dot{c}^{\dagger}_{\da}(\y) + \dot{c}_{\da}(\y) c_{\ua}(\x) + c_{\da}(\y) \dot{c}_{\ua}(\x) \right]  \nn 
 & & - A \left[ e^{ -i \Q\cdot\x} \left( \dot{c}^{\dagger}_{\ua}(\x) c_{\da}(\x) 
+ c^{\dagger}_{\ua}(\x) \dot{c}_{\da}(\x) \right) + e^{i \Q\cdot \x} \left( 
\dot{c}^{\dagger}_{\da}(\x) c_{\ua}(\x) + c^{\dagger}_{\da}(\x) 
\dot{c}_{\ua}(\x) \right) \right].
 \label{hDot}
\eeq
This expression can be simplified using the equations of motion of the 
fermionic operators,
\beq
i \, \dot{c}_{\alpha} = \left[ c_{\alpha}, H \right],
\eeq
yielding
\beq
i \, \dot{c}_{\ua}(\x) &=& - \frac{1}{2m}\nabla^2 c_{\ua}(\x) + \int d\y \, \Delta(\x - \y)  c^{\dagger}_{\da}(\y) - A \, e^{-i \Q \cdot \x} c_{\da}(\x) \nn 
i \, \dot{c}_{\da}(\x) &=& - \frac{1}{2m}\nabla^2 c_{\da}(\x) - \int d\y \, \Delta(\y - \x)  c^{\dagger}_{\ua}(\y) - A \, e^{i \Q \cdot \x} c_{\ua}(\x) 
\label{eom}
\eeq
for the above Hamiltonian. We can re-write the first term in Eq.~(\ref{hDot}) in 
the following convenient way:
\beq
\frac{1}{2m} \left[  \nabla \dot{c}^{\dagger}_{\alpha}(\x). \nabla c_{\alpha}(\x) +  \nabla c^{\dagger}_{\alpha}(\x). \nabla \dot{c}_{\alpha}(\x)  \right]  &=&  \frac{1}{2m}  \nabla \cdot \left[  \dot{c}^{\dagger}_{\alpha}(\x) \nabla c_{\alpha}(\x) + \nabla c^{\dagger}_{\alpha}(\x)  \dot{c}_{\alpha} \right] + \nn
&& \dot{c}^{\dagger}_{\alpha}(\x)  \left[ - \frac{1}{2m}\nabla^2 c_{\alpha}(\x)  \right] +  \left[ - \frac{1}{2m}\nabla^2 c^{\dagger}_{\alpha}(\x)  \right] \dot{c}_{\alpha}(\x)  \nn
\label{bypartsEq}
\eeq
Replacing the terms with Laplacians using the equation of motion, 
Eq.~\eqref{eom}, we find that fermion bilinears with two time derivatives 
cancel, and obtain for the second term in Eq.~\eqref{bypartsEq}
\beq
&& \dot{c}^{\dagger}_{\alpha}(\x)  \left[ - \frac{1}{2m}\nabla^2 c_{\alpha}(\x)  \right] +  \left[ - \frac{1}{2m}\nabla^2 c^{\dagger}_{\alpha}(\x)  \right] \dot{c}_{\alpha}(\x) =  \nn
&& - \int d\y \, \Delta(\x - \y) \left[ \dot{c}^{\dagger}_{\ua}(\x) c^{\dagger}_{\da}(\y) + c_{\da}(\y) \dot{c}_{\ua}(\x) \right] + \int d\y \, \Delta(\y - \x) \left[ \dot{c}^{\dagger}_{\da}(\x) c^{\dagger}_{\ua}(\y) + c_{\ua}(\y) \dot{c}_{\da}(\x) \right] \nn
&& +  A \,  e^{i \Q \cdot \x} \left[ \dot{c}^{\dagger}_{\da}(\x) c_{\ua}(\x) + 
c^{\dagger}_{\da}(\x) \dot{c}_{\ua}(\x) \right] + A \, e^{- i \Q \cdot \x} 
\left[ \dot{c}^{\dagger}_{\ua}(\x) c_{\da}(\x) +  c^{\dagger}_{\ua}(\x) 
\dot{c}_{\da}(\x) \right].
\label{bypartsEq2}
\eeq
Substituting the results from Eq.~(\ref{bypartsEq}) and Eq.~(\ref{bypartsEq2}) 
in Eq.~(\ref{hDot}), we find that several terms, including the terms 
proportional to the antiferromagnetic order parameter $A$, cancel. Therefore,
we can re-write Eq.~(\ref{hDot}) as:
\beq
\dot{h}(\x) &=&  \frac{1}{2m}  \nabla \cdot \left[  \dot{c}^{\dagger}_{\alpha}(\x) \nabla c_{\alpha}(\x) + \nabla c^{\dagger}_{\alpha}(\x)  \dot{c}_{\alpha} \right]  \nn 
& & + \int d\y \, \Delta(\x -\y) \left[ c^{\dagger}_{\ua}(\x) \dot{c}^{\dagger}_{\da}(\y) + \dot{c}^{\dagger}_{\da}(\x) c^{\dagger}_{\ua}(\y) + \dot{c}_{\da}(\y) c_{\ua}(\x) + c_{\ua}(\y) \dot{c}_{\da}(\x) \right]
\eeq
where we have used that $\Delta(\x - \y) = \Delta(\y - \x)$ for the $d$-wave superconductors we are interested in. Note that the first-term is already written as a divergence, so we already have 
\beq
\j_1^{Q}(\x, t) = - \frac{1}{2m} \left(  \dot{c}^{\dagger}_{\alpha}(\x) \nabla c_{\alpha}(\x) + \nabla c^{\dagger}_{\alpha}(\x)  \dot{c}_{\alpha} \right)
\label{j1Q}
\eeq
We only need to recast the second term as a divergence to find the expression for the thermal current operator. To do so, we consider the space-time Fourier transform of the second term. We set the following convention for the Fourier transform:
\beq
\j^{Q}(\x, t) = \frac{1}{V} \sum_{\q, \Omega} e^{i (\q \cdot \x - \Omega t)} \j^{Q}(\q, \Omega), \\
c_{\alpha}(\x, t) = \frac{1}{\sqrt{V}} \sum_{\k, \omega} e^{i (\k \cdot \x - \omega t)} c_{\alpha}(\k , \omega)
\eeq
Some algebra yields 
\beq
-\nabla \cdot \j^{Q}_2(\x, t) = \frac{1}{V} \sum_{\q, \Omega} e^{-i (\q \cdot \x - \Omega t)} & \bigg( \sum_{\k, \omega} ( \Delta_{\k} - \Delta_{\k  + \q }) \big[ i (\omega+ \Omega) c^{\dagger}_{\ua}(\k , \omega ) c^{\dagger}_{\da}(-\k - \q, -\omega - \Omega) \nn 
& +  i \omega c_{\da}(-\k, -\omega ) c_{\ua}(\k +  \q, \omega + \Omega) \big] \bigg)
\eeq
for the second contribution to the heat current operator. In the limit $\q 
\rightarrow 0$, we exploit
\beq
\Delta_{\k + \q} - \Delta_{\k} \approx \q \cdot \frac{\partial \Delta_{\k}}{\partial \k} = \q \cdot \vd (\k)
\eeq
and can obtain
\beq
 \j^{Q}_2(\q \rightarrow 0, \Omega) =  \sum_{\k, \omega} \vd(\k) \bigg( (\omega+ 
\Omega) c^{\dagger}_{\ua}(\k , \omega ) c^{\dagger}_{\da}(-\k - \q, -\omega - 
\Omega) +  \omega c_{\da}(-\k, -\omega ) c_{\ua}(\k +  \q, \omega + \Omega) 
\bigg). \nn
\eeq
Computing the space-time Fourier transform of Eq.~\eqref{j1Q} and using $\vf = 
\k/m$ ($ = \partial_{\k}\xi_{\k}$ for a more general dispersion), we obtain
\beq
 \j^{Q}_1(\q, \Omega) =  \sum_{\k, \omega} \left( \omega + \frac{\Omega}{2} 
\right) \vf c^{\dagger}_{\alpha}(\k , \omega)  c_{\alpha}(\k + \q, \omega + 
\Omega).
 \eeq
The total thermal current operator is thus given by
\beq
\label{jQ}
 \j^{Q}(\q \rightarrow 0, \Omega) =  \j^{Q}_1(\q \rightarrow 0, \Omega) +  
\j^{Q}_2(\q \rightarrow 0, \Omega).
\eeq

The thermal current operator does not depend on the spiral order amplitude $A$. 
One way to understand this result is to think about conductivities in terms of 
generalized velocities multiplied by occupation numbers. In our case, both the 
Fermi velocity and the gap velocity appear as the dispersion $\xi_{\k}$ and the 
gap $\Delta_{\k}$ are both momentum-dependent. However, the amplitude $A$ does 
not depend on momentum, i.e, $\partial_{\k}A = 0$, and it thus does not appear in 
the expression for the thermal current. This is similar to the absence 
of spatially uniform order parameters in thermal current operators in previous studies, e.g, for 
the s-wave on-site CDW order studied in 
Ref.~\onlinecite{DurstSachdev_PRB2009}, or the s-wave 
superconductivity in Ref.~\onlinecite{AmbegaokarGriffin} where the gap-velocity 
is zero.

\subsection{Green's function}
In momentum space, the Hamiltonian  for the clean system ($H_{dis} = 0$) can be 
written as $H = \sum_{\k}' \Psi^{\dagger}_{\k} h(\k) \Psi_{\k}$, where 
$\Psi_{\k}$ is a 2 or 4 component Nambu spinor which depends on the particular 
regime we are considering, and $\sum_{\k}' $ corresponds to the momentum sum 
over an appropriately reduced Brillouin zone (BZ). The bare Matsubara 
Green's function in the Nambu basis described above is given by
\beq
G_{0}(\k, i \omega_n) = \left( i \omega_n - h_{\k} \right)^{-1}.
\eeq
We add impurity scattering through a self-energy, which is a $2 \times 2$ or $4 
\times 4$ matrix $\hat{\Sigma}(i \omega_n)$ in Nambu space in full generality. 
Here, we only consider the scalar term for simplicity, which allows us to write 
down the dressed Green's function in terms of the bare one as follows
\begin{equation}
\begin{split}
G^{-1}(\k, i\omega_n) &= [G_{0}(\k, i\omega_n)]^{-1} - \hat{\Sigma}(i \omega_n) 
\approx [G_{0}(\k, i \omega_n)]^{-1} - \Sigma(i \omega_n)  \mathbb{I}_{4 \times 
4}\\
 &= G_{0}(\k, i \omega_n -  \Sigma(i \omega_n) )
\label{eq:AddingDisorder}
\end{split}
\end{equation}
For the computation of dc conductivities at low temperatures, we need the 
imaginary part of the retarded Green's function, $\operatorname{Im} G_{R}(\k, 
\omega)$, for $\omega \rightarrow 0$. $G_{R}(\k, \omega)$ is obtained by 
analytic continuation of the Matsubara Green's function, $G_{R}(\k, \omega) = 
G(\k, i \omega_n \rightarrow \omega + i 0^+)$. In the low-energy limit, the 
retarded self-energy from impurity scattering can be approximated as 
$\Sigma_R(0) = - i \Gamma$, where $\Gamma$ is the disorder-induced scattering 
rate.

\subsection{Kubo formula for thermal conductivity}
\label{Kubo}

\begin{figure}
\includegraphics[scale=0.3]{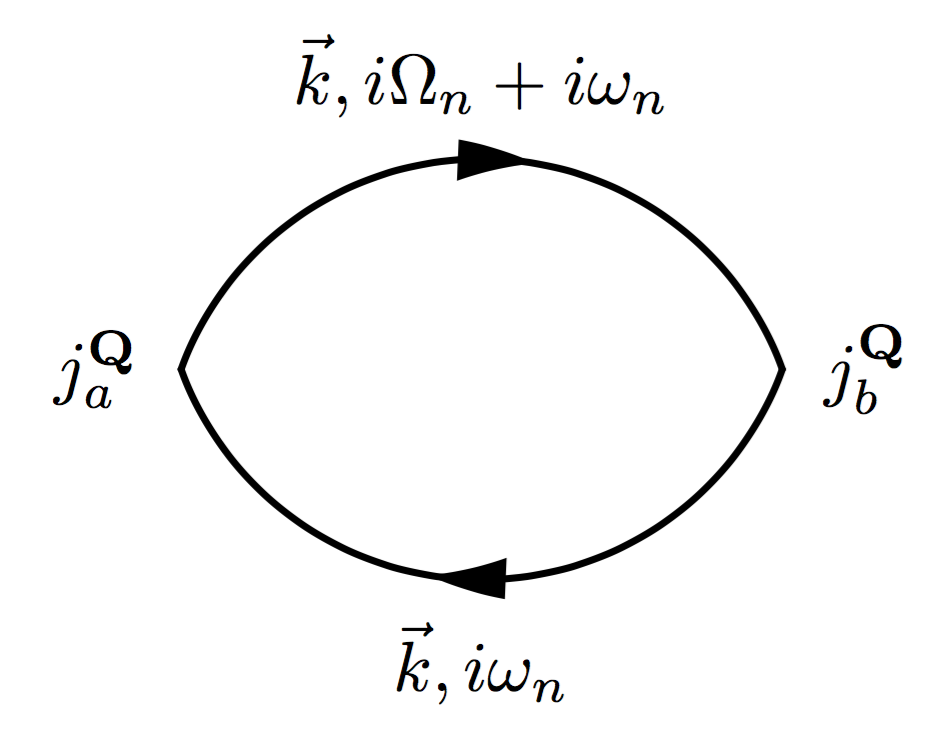}
\caption{The bare fermion bubble required to be evaluated for $\kappa_{ab}$}
\label{loop}
\end{figure}

In terms of the Nambu spinor $\Psi_\k$, the thermal current operator is given by
\beq
 \j^{Q}(\q \rightarrow 0, \Omega) =  \sum_{\k ,\omega}^{\prime}  \left( \omega + 
\frac{\Omega}{2} \right) \Psi^{\dagger}_{k} \mathbf{V}_{\k} \psi_{k + q},
 \eeq
where $\mathbf{V}_{\k}$ is a generalized velocity matrix which will be 
appropriately defined in the different scenarios. Following Ref. 
\onlinecite{DurstLee_PRB2000,DurstSachdev_PRB2009}, we can calculate the thermal 
conductivity via the Kubo formula,
\beq
\frac{\overset{\leftrightarrow}{\kappa} (\Omega,T)}{T} = - \underset{\Omega 
\rightarrow 0}{\text{lim}} \frac{\text{Im } \left[ 
\overset{\leftrightarrow}{\Pi^{R}_{\kappa}}(\Omega) \right]}{T^2 \Omega}.
\label{KappaEq1}
\eeq
$\Pi^{R}_{\kappa}(\Omega)$ is the retarded current-current correlation function 
for the thermal current, which is obtained from the Matsubara correlation 
function via analytic continuation,
\beq
\Pi^{R}_{\kappa}(\Omega) = \Pi_{\kappa}(i \Omega_n \rightarrow \Omega + i 0^+).
\eeq
In this study, we neglect vertex corrections to conductivities, so that the 
evaluation of the conductivity reduces to the evaluation of the one-loop 
contribution to the current-current correlation function. It is shown 
diagrammatically in Fig.~\ref{loop}, and is given by
\beq
\Pi_{\kappa}(i \Omega_n) = \frac{1}{ \beta} \sum_{i \omega_n, \k}' \left( i 
\omega_n+ \frac{i \Omega_n}{2} \right)^2 \text{Tr}\left[ G(\k, 
i\omega_n)\mathbf{V}_{\k} G(\k, i\omega_n + i \Omega_n)\mathbf{V}_{\k}\right],
\label{thPol}
\eeq
where $\beta = (k_B T)^{-1}$, and and the momentum sum is over the reduced 
Brillouin zone.
For the evaluation of the conductivity, it is useful to express the Green's 
function in terms of the spectral representation (for associated subtleties 
which are irrelevant for us because our order parameters are real, see the 
appendix of Ref.~\onlinecite{DurstSachdev_PRB2009}),
\beq
G(\k, i\omega_n) = \int_{-\infty}^{\infty} d\omega_1 \; \frac{A(\k, 
\omega_1)}{i\omega_n - \omega_1}, \text{ where } A(\k, \omega_1) = - 
\frac{1}{\pi} G^{\prime \prime}_{ret}(\k, \omega_1).
\eeq
Plugging this into Eq.~(\ref{thPol}), we find
\beq
\Pi_{\kappa}(i \Omega_n)  =   \int^{'} \frac{d^2k}{(2\pi)^2} \int d\omega_1 \int 
d\omega_2 \, S(i \Omega_n) \, \text{Tr}\left[ A(\k, \omega_1)\mathbf{V}_{\k} 
A(\k, \omega_2)\mathbf{V}_{\k}\right],
\eeq
where 
\beq
S(i \Omega_n) = \frac{1}{\beta} \sum_{i \omega_n} \left( i \omega_n+ \frac{i 
\Omega_n}{2} \right)^2 \frac{1}{i \omega_n - \omega_1} \frac{1}{i \omega_n + i 
\Omega_n - \omega_2}.
\label{eq:S_kappa_sloppy}
\eeq
The apparent divergence of the Matsubara sum in Eq.~\eqref{eq:S_kappa_sloppy} 
is a consequence of the improper treatment of time ordering and time 
derivatives, which do not commute. A more careful 
treatment~\cite{AmbegaokarGriffin} shows that these issues can safely be 
ignored and Eq.~\eqref{eq:S_kappa_sloppy} yields
\beq
S_{ret}(\Omega) = S(i \Omega \rightarrow \Omega + i 0^+) &=& \frac{\left( 
\omega_1 + \frac{\Omega}{2} \right)^2 n_F(\omega_1) - \left( \omega_2 - 
\frac{\Omega}{2} \right)^2 n_F(\omega_2)}{\omega_1 - \omega_2 + \Omega + i 0^+}.
\eeq
Exploiting this result, we obtain
\beq
\text{Im} \left[ \overset{\leftrightarrow}{\Pi^{R}_{\kappa}}(\Omega) \right] =  \int^{'} \frac{d^2k}{4\pi} \int d\omega \left( \omega + \frac{\Omega}{2} \right)^2 (n_F(\omega + \Omega) - n_F(\omega)) \text{Tr}\left[ A(\k, \omega)\mathbf{V}_{\k} A(\k, \omega + \Omega)\mathbf{V}_{\k}\right] \nn
\eeq
for the imaginary part of the retarded polarization bubble. The real part of 
the polarization bubble can in principle be calculated using a Kramers-Kronig 
transformation, but is not required for the computation of the dc thermal 
conductivity.  

In the static limit, where $\Omega \rightarrow 0$, we can replace $(n_F(\omega + 
\Omega) - n_F(\omega)) \rightarrow \Omega \, n_F^{\prime}(\omega)$, canceling the 
factor of $\Omega$ in the denominator of Eq.~(\ref{KappaEq1}). Subsequently, we 
can set $\Omega = 0$ everywhere else and obtain in this limit
\beq
\frac{\overset{\leftrightarrow}{\kappa} (\Omega \rightarrow 0,T)}{T} =  \int^{'} 
\frac{d^2k}{4\pi} \int d\omega \left( \frac{ \omega}{T} \right)^2 \left( - 
n_F^{\prime}(\omega) \right) \text{Tr}\left[ A(\k, \omega)\mathbf{V}_{\k} A(\k, 
\omega)\mathbf{V}_{\k}\right].
\eeq
In the scenarios we are interested in, the relevant energy scales $T$ and 
$\Gamma$ are all much smaller than the Fermi energy $E_F$. 
As the derivative of the Fermi function is strongly peaked at $\omega = 0$, for 
low $T \ll \Gamma \ll E_F$, we can set $\omega = 0$ in the spectral functions, and 
evaluate the frequency integral analytically, obtaining
\beq
\int_{-\infty}^{\infty} d\omega \left( \frac{ \omega}{T} \right)^2 \left( - 
n_F^{\prime}(\omega) \right)  = \frac{\pi^2 k_B^2}{3}
\eeq
In this limit, the conductivity takes the form:
\beq
\frac{\overset{\leftrightarrow}{\kappa} (\Omega \rightarrow 0,T)}{T} = \frac{k_B^2}{3}  \int^{'} \frac{d^2k}{4\pi} \text{Tr}\left[ G_R^{\prime \prime}(\k, 0)\mathbf{V}_{\k} G_R^{\prime \prime}(\k, 0)\mathbf{V}_{\k}\right]
\label{eq:KappaGen}
\eeq
where $G_R^{\prime \prime}(\k, 0)$ is the imaginary part of the retarded Green's 
function and the momentum integral is over the (reduced) Brillouin zone. For 
arbitrary disorder strength, this expression is difficult to 
evaluate analytically. In certain limits, we can make analytic progress and 
determine for example whether the Wiedemann-Franz law is satisfied. These 
analytic calculations will be complemented with numerical results.

\section{Antiferromagnetic metal}
\label{AFM}
\subsection{Thermal conductivity in the spiral and N\'{e}el states}
In this section, we focus on the dirty limit, where the disorder scattering 
strength is much stronger than the superconducting order, \ie, the regime where 
$\Gamma \gg \Delta_0$. In this limit, we can neglect superconductivity 
entirely, and therefore the problem reduces to the computation of the thermal 
conductivity in an antiferromagnetic metal. We proceed as described in 
Sec.~\ref{formalism}.

\begin{figure}
	\centering
	\begin{subfigure}[c]{0.35\linewidth}
		\includegraphics[width=\linewidth]{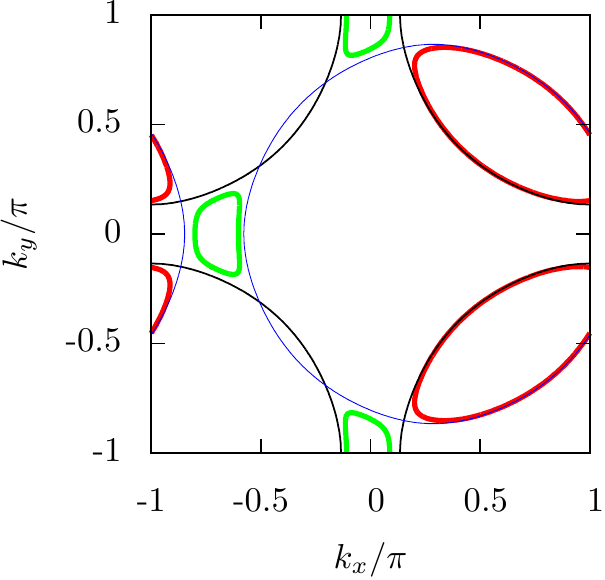}
	\end{subfigure}
	\quad
\begin{subfigure}[c]{0.35\linewidth}
\includegraphics[width=\linewidth]{%
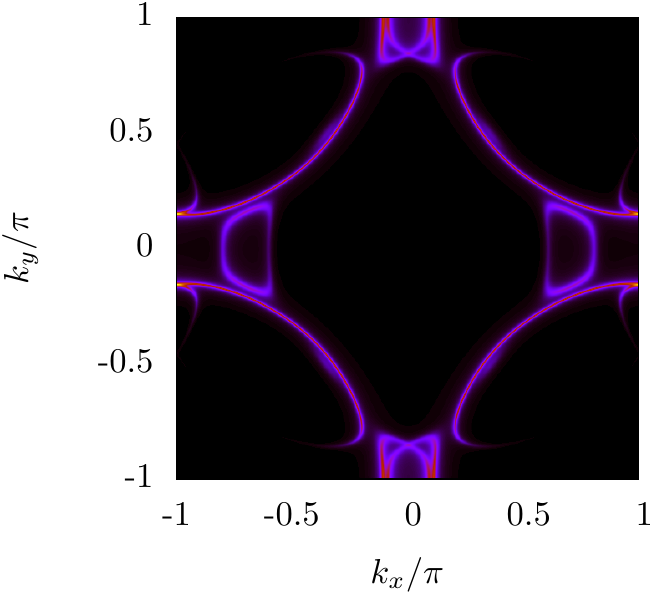}
	\end{subfigure}
\begin{subfigure}[c]{0.35\linewidth}
		\includegraphics[width=\linewidth]{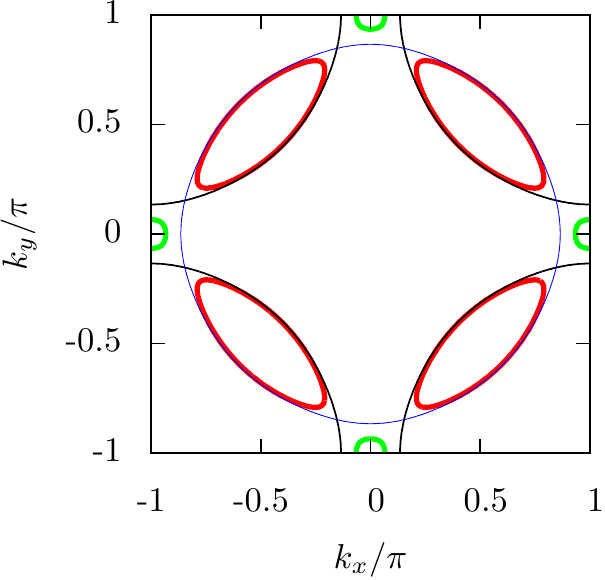}
	\end{subfigure}
	\quad
\begin{subfigure}[c]{0.35\linewidth}
		
\includegraphics[width=\linewidth]{%
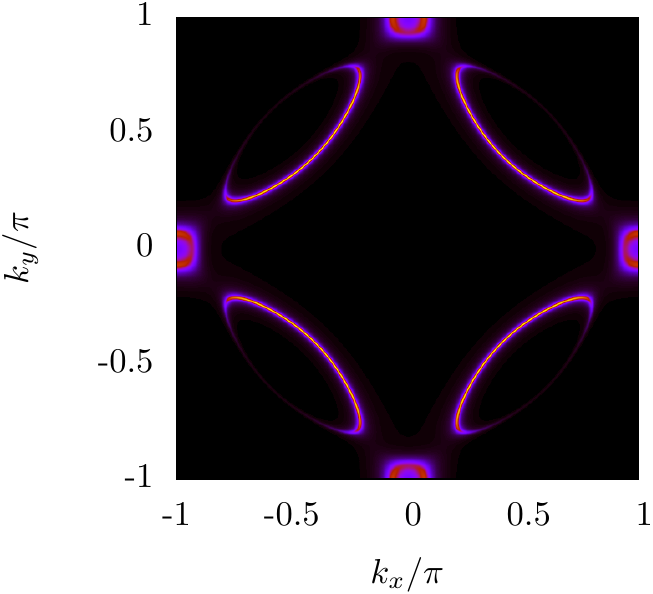}
	\end{subfigure}
	\caption{Quasi-particle Fermi surface (left) and spectral function (right) of 
spiral (top row) and commensurate (bottom row) antiferromagnetic states with 
$t' = -0.35$, $A = 0.267$, $p = 0.152$ 
and $\eta \approx p$ ($\eta = 0$) for the spiral (commensurate) AF state. In 
the left figures, hole and electron pockets are marked in red and green, 
respectively. The thin black and blue lines are the original and $\boldsymbol 
Q$-shifted Fermi surfaces.}
\label{fig:AF_FS_A}
\end{figure}

An antiferromagnetic state with ordering wave vector $\Q$ can be described by 
the Hamiltonian
\beq
H_{afm} &=& \sum_{\k, \sigma= \uparrow,\downarrow} \xi_{\k} c^{\dagger}_{\k 
\sigma} c_{\k \alpha} - A \sum_{\bs k} \left( c^{\dagger}_{\k \ua} c_{\k + \Q 
\da}  + c^{\dagger}_{\k + \Q \da} c_{\k \ua} \right) \nn
& = &  \sum_{\k}  \Psi^{\dagger}_{\k} h_{\k} \Psi_{\k} \text{ with }
h(\k) = \begin{pmatrix}  \xi_{\k} & -A \\
-A & - \xi_{\k} 
\end{pmatrix}
\text{ and } \Psi_{\k} = \begin{pmatrix}
c_{\k \uparrow} \\
c_{\k + \Q \downarrow} 
\end{pmatrix} \nn
\label{HiAF}
\eeq 
 The Green's function is given by
\beq
G_{0}(\k, i \omega_n) = (i \omega_n - h_{\k})^{-1} = \frac{1}{(i \omega_n - 
E_{+ \k})(i \omega_n - E_{- \k})} \begin{pmatrix} i \omega_n - \xi_{\k} & A \\
A & i \omega_n - \xi_{\k + \Q} 
\end{pmatrix}
\label{eq:G_iAF_metal}
\eeq
where
\begin{equation}
E_{\pm,\k} = \frac{\xi_{\k} + \xi_{\k + \Q} }{2} \pm \sqrt{ 
\left( \frac{\xi_{\k} - \xi_{\k + \Q}}{2} \right)^2 + A^2 }
\end{equation}
are the two reconstructed bands. An example for the quasi-particle Fermi 
surface and spectral function of this Hamiltonian is shown in 
Fig.~\ref{fig:AF_FS_A}~\cite{Eberlein2016}.

For the transport calculation, disorder is added to the Green's function as 
described in Eq~\eqref{eq:AddingDisorder}. In the spiral state, the momenta and 
the spins of the quasiparticles are tied together. This may in general lead to 
a momentum and spin dependence of the scattering rate even for potential 
disorder. However, since each state can get scattered to any 
other state on the Fermi surface by repeated scattering, we assume that the 
averaged scattering cross-section is roughly the same for any given momenta on 
the Fermi surface. Therefore, we use a simple retarded self-energy 
$\Sigma_{R}(\omega \rightarrow 0) = - i \Gamma$ to account for the 
broadening of the quasiparticle spectrum near the Fermi surface. 

The heat conductivity is then evaluated using Eq.~\eqref{eq:KappaGen}, using 
the imaginary part of Eq.~\eqref{eq:G_iAF_metal} after continuation to the real 
frequency axis, $i \omega_n \rightarrow \omega + i \delta$,
\begin{equation}
\text{Im}\left[ G^{R} (\k, \omega \rightarrow 0) \right] =
\frac{\Gamma}{G_{den}} \left[ - \left( (E_{+ \k}^2 +  E_{- \k}^2)/2 + 
\Gamma^2 \right) \tau_{0} + \Gamma (E_{+ \k} + E_{- \k}) \left( A \, \tau_{1}- 
\frac{(\xi_{\k} - \xi_{\k+\Q})}{2}\, \tau_{3}  \right)  \right],
\label{ImGRiAF}
\end{equation}
where $G_{den} = (E_{+ \k} E_{- \k} - \Gamma^2 )^2 + \Gamma^2 
\left( E_{+ \k} + E_{- \k} \right)^2$, $\tau_i$ are the Pauli matrices; and the velocity matrix,
\begin{equation}
	\mathbf{V}_{\k} = \begin{pmatrix}
 \mathbf{v}(\k)  & 0 \\
 0 & \mathbf{v}(\k + \Q) 
 \end{pmatrix}
\end{equation}
with appropriate $\Q$ for the antiferromagnetic state under consideration. This yields the 
thermal current operator in Eq.~\eqref{jQ} with $\j^Q_2 = 0$. Note that the 
momentum integral in Eq.~\eqref{eq:KappaGen} is over the whole Brillouin zone 
for a spiral antiferromagnetic metal.

This expression needs to be evaluated numerically, but we can simplify it to 
some extent in the limit where the disorder strength $\Gamma$ is smaller than 
all the other relevant energy scales, the amplitude of the antiferromagnetic 
order $A$ and the Fermi energy $E_F$ (but it is still larger than $\Delta_d$). The final expression, which we do not state here, involves an integral along the Fermi pockets, and is inversely proportional to the scattering rate $\Gamma$.

\subsection{Electrical conductivity and the Wiedemann-Franz Law}
\label{afWF}
In this section we evaluate the electrical conductivity for the dirty metal with 
N\'{e}el or spiral order, and show that the Wiedemann Franz law for transport 
holds. 
The electric current operator is given by:
\beq
 \j^{e}(\q \rightarrow 0, \Omega) &=& \sum_{\k ,\omega, \sigma}  \frac{\partial 
\xi_{\k}}{\partial \k} \; c^{\dagger}_{ \sigma}(\k, \omega) 
c_{\sigma}(\k,\omega)  =  \sum_{\k ,\omega}  \Psi^{\dagger}_{k} \mathbf{V}_{\k} 
\Psi_{k + q}, \nn \text{ where }
\mathbf{V}_{\k} &=& \begin{pmatrix}
 \mathbf{v}(\k)  & 0 \\
 0 & \mathbf{v}(\k + \Q) 
 \end{pmatrix}
\eeq
The Kubo formula for the electrical conductivity is given by:
\beq
\overset{\leftrightarrow}{\sigma} (\Omega,T) = - \underset{\Omega \rightarrow 
0}{\text{lim}} \frac{\text{Im } \left[ 
\overset{\leftrightarrow}{\Pi^{R}_{e}}(\Omega) \right]}{ \Omega}
\label{SigmaEq1}
\eeq
where $\Pi^{R}_{e}(\Omega)$ is the retarded current-current correlation function 
for the electrical current, obtained via analytic continuation from the 
Matsubara correlation:
\beq
\Pi^{R}_{e}(\Omega) = \Pi_{e}(i \Omega_n \rightarrow \Omega + i 0^+)
\eeq
Neglecting vertex corrections, we evaluate the bare-bubble contribution to the 
current-current correlator:
\beq
\Pi_{e}(i \Omega_n) = \frac{1}{\beta} \sum_{i \omega_n, \k} \text{Tr}\left[ 
G(\k, i\omega_n)\mathbf{V}_{\k} G(\k, i\omega_n + i 
\Omega_n)\mathbf{V}_{\k}\right]
\label{elPol}
\eeq
Using the spectral representation of the Green's function in Eq.~(\ref{elPol}), 
we find:
\beq
\Pi_{e}(i \Omega_n)  =  \int \frac{d^2k}{(2\pi)^2} \int d\omega_1 \int d\omega_2 
\, S(i \Omega_n) \, \text{Tr}\left[ A(\k, \omega_1)\mathbf{V}_{\k} A(\k, 
\omega_2)\mathbf{V}_{\k}\right]
\eeq
where 
\beq
S(i \Omega_n) = \frac{1}{\beta} \sum_{i \omega_n} \frac{1}{i \omega_n - 
\omega_1} \frac{1}{i \omega_n + i \Omega_n - \omega_2}
\eeq
The Matsubara sum is convergent, and evaluates to:
\beq
S_{ret}(\Omega) = S(i \Omega \rightarrow \Omega + i 0^+) &=& \frac{ 
n_F(\omega_1) - n_F(\omega_2)}{\omega_1 - \omega_2 + \Omega + i 0^+} \nn 
\implies \text{Im}\left[  S_{ret}(\Omega) )  \right] & = & \pi (n_F(\omega_1 + 
\Omega) - n_F(\omega_1)) \, \delta(\omega_1 - \omega_2 + \Omega) 
\eeq
The imaginary part of retarded polarization bubble is therefore given by:
\beq
\text{Im} \left[ \overset{\leftrightarrow}{\Pi^{R}_{e}}(\Omega) \right] =  \int 
\frac{d^2k}{4 \pi} \int d\omega  (n_F(\omega + \Omega) - n_F(\omega)) 
\text{Tr}\left[ A(\k, \omega)\mathbf{V}_{\k} A(\k, \omega + 
\Omega)\mathbf{V}_{\k}\right] 
\eeq
Assuming that $T \rightarrow 0$, we can approximate the derivative of the Fermi 
function by a delta function as $\Omega \rightarrow 0$ in the dc limit:
\beq
(n_F(\omega + \Omega) - n_F(\omega))/\Omega \approx \delta(\omega)
\eeq
Using this to evaluate the $\omega$ integral, we end up with the following 
expression for the electrical conductivity:
\beq
\overset{\leftrightarrow}{\sigma} (\Omega \rightarrow 0,T \rightarrow 0) = 
\frac{e^2}{\pi^2} \int \frac{d^2k}{4\pi} \text{Tr}\left[ G_R^{\prime \prime}(\k, 
0)\mathbf{V}_{\k} G_R^{\prime \prime}(\k, 0)\mathbf{V}_{\k}\right]
\label{Sigma_iAF}
\eeq
Comparing Eq.~(\ref{Sigma_iAF}) with Eq.~\eqref{eq:KappaGen}, we find that 
the Wiedemann-Franz law is obeyed, as one would expect for a quasiparticle 
Fermi surface with constant scattering lifetime:
\beq
\frac{\kappa}{\sigma T} = \frac{\pi^2 k_B^2}{3 e^2}
\eeq

\subsection{Numerical results for antiferromagnetic metals}

\begin{figure}
\centering
	\includegraphics[width=0.48\linewidth]{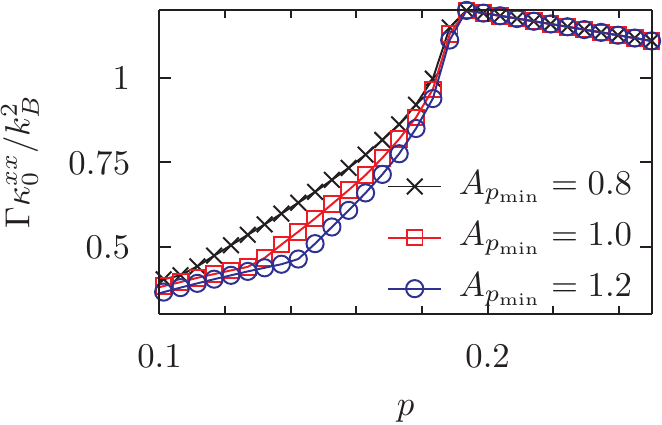}
	\hspace{0.02\linewidth}
	\includegraphics[width=0.48\linewidth]{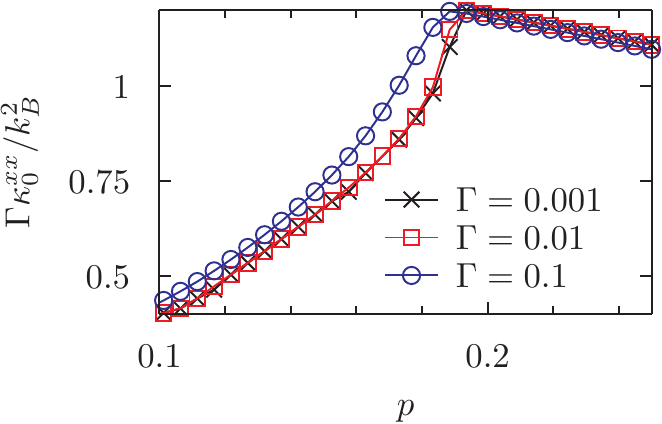}
\caption{Influence of the size of the spiral antiferromagnetic gap $A$ (left) 
and the scattering rate $\Gamma$ (right) on the heat conductivity for $t' = 
-0.35$.  and $\Gamma = 0.01$. Antiferromagnetism 
disappears at $p^\ast = 0.19$. In the left plot we use 
$\Gamma = 0.01$ and in the right plot the gap at $p = 0.05$ 
is set to $A_{p_\text{min}} = 0.8$.}
\label{fig:Influence_iAF}
\end{figure}
In the following we discuss numerical results for
\begin{equation}
	\kappa_0 = \lim_{T \rightarrow 0} \frac{\kappa_{xx}(\Omega\rightarrow 0, 
T)}{T}
\end{equation}
in the presence of (in-) commensurate antiferromagnetic order. The latter is 
described by an antiferromagnetic gap with the phenomenological doping dependence
\begin{equation}
	A(p) = A_0 (p^* - p) \Theta(p^* - p),
\end{equation}
where $A_0$ is fixed by the antiferromagnetic gap $A_{p_\text{min}}$ at the 
smallest doping considered, $p_\text{min} = 0.05$ and $p^* = 0.19$ is the critical 
doping beyond which antiferromagnetic order disappears. In this 
section, we set $t' = -0.35$. In Fig.~\ref{fig:Influence_iAF} we show the 
influence of the size of the 
antiferromagnetic gap on the heat conductivity. Similarly to the Hall 
coefficient near an antiferromagnetic phase 
transition~\cite{Storey2016,Eberlein2016}, the magnitude of the spiral order 
parameter mostly influences the width of the crossover region, in which electron 
and hole pockets coexist. With increasing strength of antiferromagnetic order, 
the transition region shrinks. In Fig.~\ref{fig:Influence_iAF}, we 
show the influence of the size of $\Gamma$ on the doping dependence of the heat 
conductivity. We plot $\Gamma \kappa_0$ for better comparison. For the smallest 
value, $\Gamma = 0.001$, the results are indistinguishable from those obtained 
in a relaxation time approximation in which $\Gamma \rightarrow 0$ is assumed, 
as employed in Ref.~\onlinecite{Eberlein2016}. The doping dependence of 
$\Gamma \kappa_0$ is only altered at large values of $\Gamma$, as shown in 
Fig.~\ref{fig:Influence_iAF}.

\begin{figure}
\centering
	\includegraphics[width=0.5\linewidth]{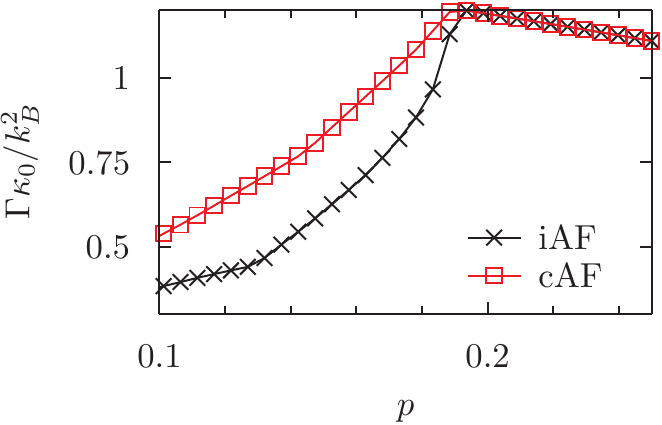}
\caption{Comparison of the doping dependence of the heat conductivity for 
commensurate (cAF) and incommensurate (iAF) antiferromagnetism for $p < p^\ast$ 
for $t' = -0.35$ and $A_{p_\text{min}} = 1.0$. Antiferromagnetism disappears at 
$p^\ast = 0.19$. We plot $\Gamma \kappa_0$ for 
better comparison.}
\label{fig:Comparison_iAF_cAF}
\end{figure}
In Fig.~\ref{fig:Comparison_iAF_cAF}, we compare the doping dependence of the 
heat conductivity for commensurate or incommensurate antiferromagnetism for $p 
< p^\ast$. In the incommensurate case, the heat conductivity drops 
significantly faster for $p < p^\ast$ than in the commensurate case. This 
difference can be understood analytically and is 
discussed in the next section.

Collignon~\etal\ found that the drop in the electrical conductivity can be 
entirely understood as a drop in the charge carrier density when assuming that
the charge carrier mobility is constant across the phase 
transition~\cite{Collignon2016}. Evidence for a constant mobility is 
found in the behavior of the Hall angle and the magnetoresistance. The 
charge carrier density can then be extracted from the heat or electrical 
conductivity via
\begin{equation}
	p_\sigma = (1 + p) \frac{\sigma(0)}{\sigma_0}
\end{equation}
where $\sigma_0$ ($\sigma(0)$) is the conductivity in the absence (presence) of 
antiferromagnetic order. In experiments, $\sigma_0$ is obtained by 
extrapolating the conductivity from high to low temperatures.

\begin{figure}
\centering
	\includegraphics[width=0.5\linewidth]{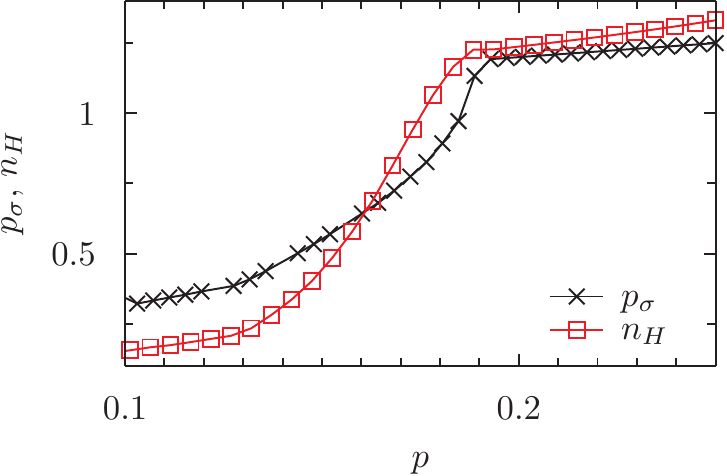}
\caption{Effective charge carrier density as extracted from the heat 
conductivity in comparison to the Hall number for a phase transition to 
incommensurate antiferromagnetic order. Parameters are $t' = -0.35$ and 
$A_{p_\text{min}} = 1.0$. $\Gamma = 0.001$ in the computation of the 
conductivity. The Hall number was determined in the relaxation time 
approximation as described in Ref.~\cite{Eberlein2016}.}
\label{fig:Comparison_pSigma_nH}
\end{figure}
\begin{figure}
\centering
	\includegraphics[width=0.5\linewidth]{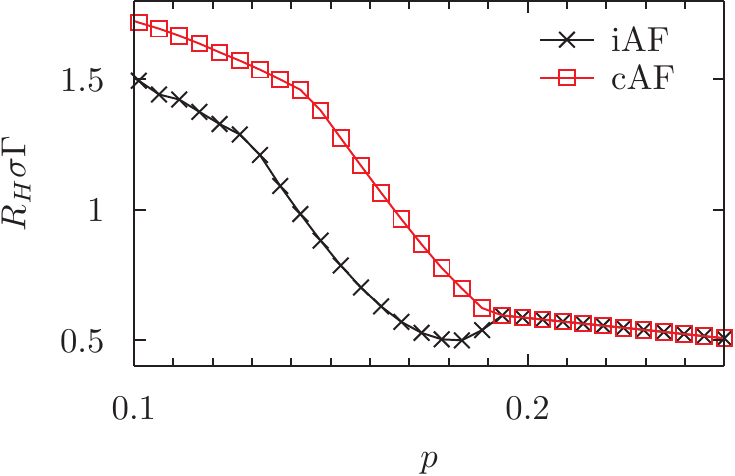}
\caption{Doping dependence of the Hall angle $R_H \sigma$, rescaled by 
$\Gamma$, across a phase 
transition from a paramagnetic to an (in-) commensurate antiferromagnetic 
metal with $A_{p_\text{min}} = 1.0$ in both cases.}
\label{fig:Plot_RH_sigma}
\end{figure}
For our calculation we assume that the disorder scattering rate is independent 
of doping and do not make assumptions on the mobility. This leads to a
behavior of $p_\sigma$ that is distinct from that of the Hall coefficient, as 
shown in Fig.~\ref{fig:Comparison_pSigma_nH}, and is connected to the fact that 
the drop in the conductivity is smaller than the drop in the Hall number. In 
Fig.~\ref{fig:Plot_RH_sigma}, we plot the Hall angle across the phase 
transition to (in-) commensurate antiferromagnetism. For commensurate 
antiferromagnetism, the Hall angle increases for $p < p^\ast$. In the 
incommensurate case, there is a slight drop of $R_H \sigma$ below the critical 
point before it starts to increase. It is interesting to compare this behavior 
to the experimental results by Collignon~\etal~\cite{Collignon2016} on 
$\textrm{Nd}$-$\textrm{LSCO}$. In Fig.~10 of their paper the Hall angle is shown to drop by a factor of three (approximately) within a small doping range. This is attributed to Fermi surface reconstruction below $p^\ast$ and appearance of electron-like pockets. In our model the relative drop is smaller, and
in Sec.~\ref{sec:DopingDependentScattering} we demonstrate how a doping dependent 
scattering rate can alleviate this discrepancy.

\subsection{Analytic understanding of the numerical results}
\label{sec:AnalyticAFMetal}
In this section, we estimate the drop of the conductivity across the phase 
transition. Our estimates are valid for the weakly disordered antiferromagnetic 
metal, or coexisting antiferromagnetism and superconductivity in the dirty 
limit ($\Delta \ll \Gamma$) and capture the change quite well. Instead 
of the thermal conductivity, we focus on the diagonal electrical conductivity 
as in this regime the Wiedemann-Franz law is obeyed.

In our mean-field model, the Fermi velocity $v_F$ is unrenormalized across the phase transition 
on most parts of the Fermi surface, except close to the points that get gapped out (near the edges of the pockets).
Therefore, the electrical conductivity can be approximated as:
\beq
\sigma_{xx} \sim \int_{\k} \left( \frac{\partial E_{\k}}{\partial k_x} \right)^2 n_F^{\prime}(E_{\k}) \approx \int_{\k} (\vf_x)^2 \delta\left(\vf(\k)\cdot(\k - \k_F)\right) \\
 \approx \left< \frac{v_{F,x}^2}{v_F} \right> \times \text{Total perimeter of 
Fermi surface/pockets}.
\label{eq:AnalyticUnderstandingApprox}
\eeq
We have also assumed that scattering is mainly due to disorder, so that the 
quasi-particle scattering rate is unchanged across the transition. 

In absence of antiferromagnetic order, we have a large hole-like Fermi surface 
of size $1+p$, which accommodates both spin species. We assume that this pocket 
is circular with radius $k_F$, so that the density of holes is given by 
(setting lattice spacing $a=1$):
\beq
2 \times \frac{\pi k_F^2}{(2\pi)^2} = 1 + p \implies k_F = \sqrt{2\pi (1 + p)}.
\eeq
This implies that the diagonal conductivity is given by (modulo constant factors):
\beq
\sigma_{Large \, FS} \sim \left< \frac{v_{F,x}^2}{v_F} \right> \times 2 \times 2 \pi k_F =  4 \pi \left< \frac{v_{F,x}^2}{v_F} \right>  \sqrt{2\pi (1 + p)}
\eeq

Antiferromagnetic order leads to reconstruction of the large Fermi surface into 
electron and hole pockets. In the following we assume that the 
antiferromagnetic order parameter is large enough to gap out the electron 
pockets. In the presence of N\'{e}el order, we have four hole pockets in the 
large 
Brillouin zone (taking spin already into account). For the spiral order there 
are only two somewhat larger hole pockets, see 
Figs.~\ref{fig:AF_FS_A}. These pockets are 
approximately elliptic, with an eccentricity of $e \approx 0.5$ in both cases. 
The area of the ellipse is given by $\pi K_1 K_2 = \pi K_1^2/2$, where 
$K_1$($K_2$) is the semi-major(minor) axis of the ellipse. The perimeter of a 
single elliptical pocket is given by
\beq
S_{ellipse} = 4 K_1 \int_{0}^{\pi/2} d\theta \sqrt{1 - e^2 \text{ sin}^2\theta} 
~ \approx 6 K_1, \text{ for } e = 0.5.
\eeq
For the N\'{e}el ordered case, we find
\beq
4 \times \frac{\pi K_1^2}{2(2\pi)^2} = p \implies K_{1} =  \sqrt{2\pi p},
\eeq
so that the diagonal conductivity is given by
\beq
\sigma_\text{N\'{e}el} \sim  \left< \frac{v_{F,x}^2}{v_F} \right> \times 4 
\times 6 K_{1} 
= 24 \left< \frac{v_{F,x}^2}{v_F} \right> \sqrt{2 \pi p}.
\eeq
For spiral order, we have
\beq
2 \times \frac{\pi K_1^2}{2(2\pi)^2} = p \implies K_{1} =  2\sqrt{\pi p}
\eeq
and can estimate the conductivity as
\beq
\sigma_{iAF} \sim  \left< \frac{v_{F,x}^2}{v_F} \right> \times 2 \times 6 K_{1} 
= 24 \left< \frac{v_{F,x}^2}{v_F} \right> \sqrt{\pi p}.
\eeq

We estimate the drop of the conductivity across the phase transition by 
comparing the results for the large Fermi surface at a doping $p_1 = 0.2$ 
with the result for the small Fermi pockets at $p_2 = 0.1$. For N\'{e}el order 
we 
find
\beq
\frac{\sigma_\text{N\'{e}el}}{\sigma_{Large \, FS}} = 
\frac{6}{\pi}\sqrt{\frac{p_2}{1+p_1}} = \frac{\sqrt{3}}{\pi} \approx 0.55,
\eeq
while for spiral order we obtain
\beq
\frac{\sigma_{iAF}}{\sigma_{Large \, FS}} = \frac{6}{\pi} \sqrt{\frac{p_2}{2(1+p_1)}} \approx 0.39
\eeq
Both of these seem to agree quite well with the numerical data, as does the 
approximation that $\sigma_\text{N\'{e}el}/\sigma_\text{iAF} = \sqrt{2}$ at the 
same doping $p_2$ after the disappearance of the electron pockets.

\section{Co-existing antiferromagnetism and superconductivity}
\label{AFMandDSC}
In this section, we discuss the thermal conductivity for co-existing 
antiferromagnetic and superconducting order. This is motivated by the fact that 
most transport experiments at low temperatures are done in the superconducting 
phase. The reason is that the experimentally accessible magnetic fields do not 
suffice to suppress superconductivity completely in most materials. Therefore 
it is interesting to ask which experimental signatures of incommensurate 
antiferromagnetic or topological order could show up in transport measurements 
in the superconducting phase. We consider both commensurate (N\'{e}el) and 
incommensurate (spiral) antiferromagnetic order. Since the formalism has a 
significant amount of overlap for these two scenarios, we combine them into a 
single section, with separate subsections where the results differ 
significantly. 

\subsection{Spectrum}
Commensurate and incommensurate antiferromagnetism coexisting with 
superconductivity can be described by the mean-field Hamiltonian in 
Eq.~\eqref{Htot}~\cite{Yamase2016}. For convenience of calculation, we re-write 
it in terms of a $4 \times 4$ Nambu notation, with
\beq
H_\text{AF+dSC} = \sum_{\k}' \Psi^{\dagger}_{\k} h(\k) \Psi_{\k}, \text{ with }
h(\k) =  \begin{pmatrix}  \xi_{\k} & \Delta_{\k} &  -A & 0 \\
\Delta_{\k} & - \xi_{\k} & 0 & A \\
-A & 0 & \xi_{\k + \Q} & -\Delta_{\k + \Q} \\
0 & A & -\Delta_{\k + \Q} & -\xi_{\k + \Q}
\end{pmatrix}
\text{ and } \Psi_{\k} = \begin{pmatrix}
c_{\k \uparrow} \\
c^{\dagger}_{-\k \downarrow}  \\
c_{\k + \Q \downarrow} \\
c^{\dagger}_{-\k - \Q \uparrow}.
\end{pmatrix} \nn
\eeq
Here we have assumed that the electron dispersion $\xi_{\k}$ is symmetric under 
spatial inversion, $\k \rightarrow - \k$. The sum over momenta $\bs 
k$ is restricted to $-\pi < k_x \leq \pi$ and $-\pi/2 < k_y \leq \pi/2$ in 
order to avoid double counting. We have checked explicitly that the mean-field 
Hamiltonian can be rewritten in the spinor notation by employing all allowed 
operations (like shifting the $x$-component of momenta, inverting momenta, but 
not shifting the $y$-component of momenta after introducing the reduced 
BZ). For general $\Q$, the eigenvalues of $h(\k)$ are given by:
\beq
E_{\pm,\k}^2 &=& \frac{1}{2} \left( 2 A^2 + \Lambda_{\k} \pm \sqrt{ \alpha_{\k}^2 + 4 A^2 \beta_{\k}} \right), \text{ where } \nn
\Lambda_{\k} &=& \Delta_{\k}^2 + \Delta_{\k + \Q}^2 + \xi_{\k}^2 + \xi_{\k + \Q}^2, \, \alpha_{\k} = \Delta_{\k}^2 - \Delta_{\k + \Q}^2 + \xi_{\k}^2 - \xi_{\k + \Q}^2, \text{ and } \nn 
\beta_{\k} &=& \left(  \Delta_{\k} +  \Delta_{\k+\Q} \right)^2 +  \left(  \xi_{\k} + \xi_{\k+ \Q} \right)^2
\label{cafEigs} 
\eeq
Setting $\Delta_{\k} = 0$, we recover the spectrum of the antiferromagnetic 
metal~\cite{Schulz1990}. For $A = 0$, we recover the spectrum of a 
superconductor $E^{0}_{\k}$ and $E^0_{\k+ \Q}$, where $E^{0}_{\k} = 
\sqrt{\xi_{\k}^2 + \Delta_{\k}^2}$. These two branches together count the states 
of the uniform superconductor in the full BZ, as there is no translation 
symmetry breaking for $A=0$.

\begin{figure}[h!]
 
\begin{subfigure}{0.45\textwidth}
\includegraphics[width=1.2\linewidth]{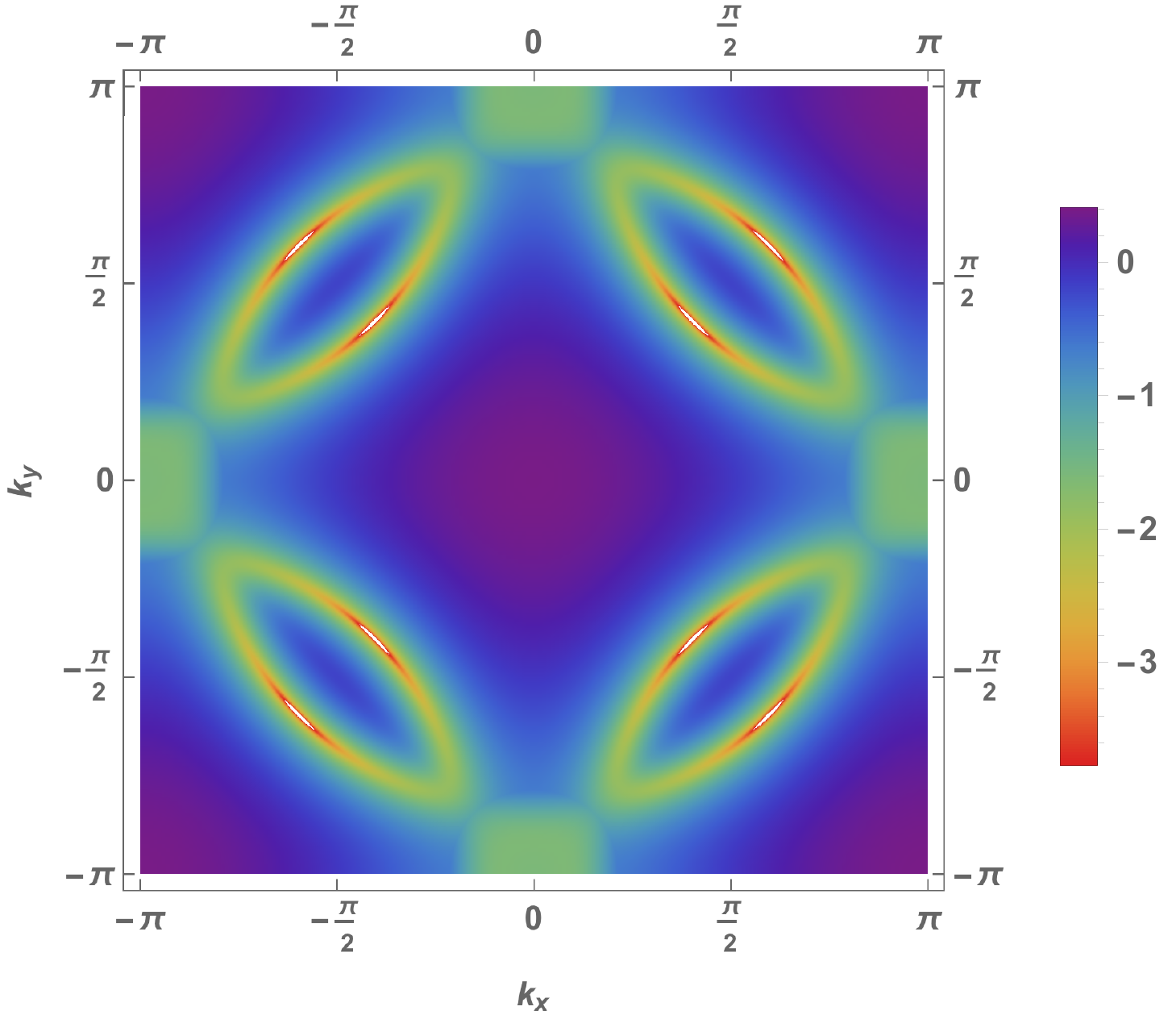} 
\caption{}
\label{fig:subim1}
\end{subfigure}
\begin{subfigure}{0.45\textwidth}
\includegraphics[width=1.2\linewidth]{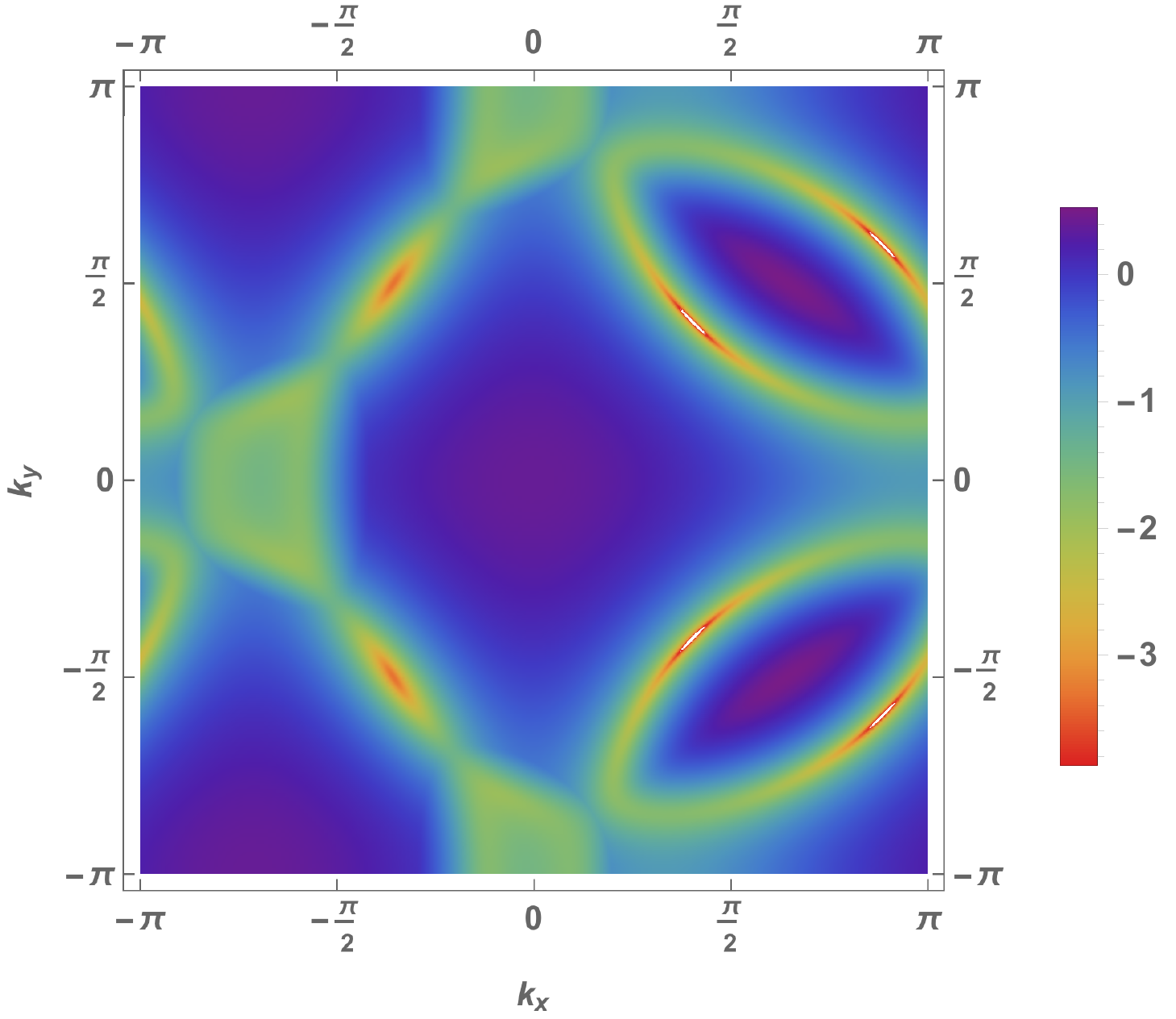}
\caption{}
\label{fig:subim2}
\end{subfigure}
\caption{Plots of logarithm of the dispersions of the lower band (a.u.) at $p = 
0.152$, for parameter values $t^{\prime} = -0.35, \mu = -1.099, \Delta_d = 0.1, 
A = 0.267$, and $\eta = 0$ for N\'{e}el and $0.1436$ for spiral order. (a) N\'{e}el order + superconductivity: Eight nodes in the extended BZ. (b) Spiral order + superconductivity: Four nodes in the extended BZ.}
\label{fig:disPlot}
\end{figure}

We choose the superconducting gap to have $d$-wave symmetry, as appropriate for 
cuprate superconductors and theoretical models of antiferromagnetism coexisting 
with superconductivity~\cite{Halboth2000,Maier2000,Lichtenstein2000,Sushkov2004,
Maier2005,Capone2006,Aichhorn2006,Khatami2008,Kancharla2008,Friederich2011,
Gull2013,Eberlein2014,Roemer2016,Yamase2016,Zheng2016}. The dispersion in 
Eq.~\eqref{cafEigs} then possesses gapless nodal excitations. In 
Fig.~\ref{fig:disPlot}, we plot the spectrum of the N\'eel and spiral states 
coexisting with superconductivity. We note that the 
N\'{e}el ordered superconductor has eight nodal points in the extended BZ, whereas 
the superconductor with spiral order has only four nodes. 


\subsection{Thermal conductivity}
The bare Matsubara Green's function is in the Nambu basis described above is given by:
\beq
G^{0}(\k, i \omega_n) = \left( i \omega_n - h_{\k} \right)^{-1}
\eeq
 
Now we add impurity contribution to the self-energy, which is a $4 \times 4$ matrix $\hat{\Sigma}(i \omega_n)$ in Nambu space in full generality. We only consider the scalar term for simplicity, which allows us to write down the dressed Green's function in terms of the bare one as follows:
\begin{equation}
\begin{split}
G^{-1}(\k, i\omega_n) &= [G^{0}(\k, i\omega_n)]^{-1} - \hat{\Sigma}(i \omega_n) 
\approx [G^{0}(\k, i \omega_n)]^{-1} - \Sigma(i \omega_n)  \mathbb{I}_{4 \times 
4}\\
 &= [G^{0}(\k, i \omega_n -  \Sigma(i \omega_n) )]^{-1}
\end{split}
\end{equation}
Following analytic continuation to real frequencies, $\Sigma_R(0) = - i 
\Gamma$, the imaginary part of the retarded Green's function in the $ \omega 
\rightarrow 0$ limit can be expressed as follows in terms of Pauli matrices 
$\tau_i$ and the $2\times2$ identity matrix $\tau_0$ (relabeling $\k $ as 1 and 
$\k + \Q$ as 2):
\beq
\text{Im}[ G_R(\k, \omega \rightarrow 0)] &=& \frac{1}{G_{den}} \begin{pmatrix}
G_a^{\prime \prime} & G_b^{\prime \prime} \\
G_c^{\prime \prime} & G_d^{\prime \prime}
\end{pmatrix}, \text{ where} \nn
G_a^{\prime \prime} & = & - \Gamma (\Gamma^2 + A^2 + \xi_2^2 + \Delta_2^2) \, \tau_0 \nn
G_b^{\prime \prime} & = &  - A \Gamma \left[ (\xi_1 + \xi_2) - (\Delta_1 + 
\Delta_2)(i \tau_2) \right] \nn
G_c^{\prime \prime} & = & - A \Gamma \left[ (\xi_1 + \xi_2) + (\Delta_1 + 
\Delta_2)(i \tau_2) \right]  \nn
G_d^{\prime \prime} & = & - \Gamma (\Gamma^2 + A^2 + \xi_1^2 + \Delta_1^2) \, \tau_0 \nn
G_{den} & = & (\Gamma^2 + A^2 + \xi_1^2 + \Delta_1^2)(\Gamma^2 + A^2 + \xi_2^2 
+ \Delta_2^2) - A^2 \left[ (\xi_1 + \xi_2)^2 + (\Delta_1 + \Delta_2)^2 \right] 
\nn
\label{ImGR} 
\eeq

Noting that $\vd(-\k) = - \vd(\k)$ and working through some algebra, we find that we can write the thermal current operator in terms of the Nambu spinor $\Psi_{\k}$ as:
\beq
 \j^{Q}(\q \rightarrow 0, \Omega) =  \sum_{\k ,\omega}^{\prime}  \left( \omega + \frac{\Omega}{2} \right) \Psi^{\dagger}_{k} \mathbf{V}_{\k} \psi_{k + q}, \text{ where } \nn
\mathbf{V}_{\k} = \begin{pmatrix}
 \vf(\k) \tau_3 + \vd(\k) \tau_1 & 0 \\
 0 &  \vf(\k + \Q) \tau_3 - \vd(\k + \Q) \tau_1
 \end{pmatrix}
\eeq
Following the procedure outlined in Sec.~\ref{Kubo}, we can evaluate the thermal conductivity by extending the summation to the full BZ with an added factor of half:
\beq
\frac{\overset{\leftrightarrow}{\kappa} (\Omega \rightarrow 0,T)}{T} = 
\frac{k_B^2}{3} \int \frac{d^2k}{8\pi} \text{Tr}\left[ G_R^{\prime \prime}(\k, 
0)\mathbf{V}_{\k} G_R^{\prime \prime}(\k, 0)\mathbf{V}_{\k}\right].
\label{KappaLimcAF}
\eeq
For arbitrary disorder strength, we evaluate this expression numerically. In 
the clean limit, it can be treated analytically so that connections with the 
universal Durst-Lee result in the absence of magnetism~\cite{DurstLee_PRB2000} 
can be drawn. This is discussed in the next section.

\subsection{Analytic expressions in the clean limit}

\subsubsection{N\'{e}el order}
\label{sec:CLcAF}
For N\'{e}el-type antiferromagnetic order coexisting with superconductivity in 
the clean limit, $\Gamma_0 \rightarrow 0$, the thermal conductivity can be 
evaluated analytically. In this case, the major contribution to the thermal 
current is carried by nodal quasiparticles, again allowing us to linearize 
the dispersion at each nodal point. We obtain
\beq
\frac{\kappa_{ii}(\Omega \rightarrow 0,T)}{T} =  \frac{k_B^2}{3 v_F v_{\Delta}} 
\bigg[ \sqrt{1 - \alpha^2} \, v_F^2 + \frac{1}{\sqrt{1 - \alpha^2}} 
v_{\Delta}^2 
\bigg] \Theta(1 - \alpha), \text{ where } \alpha = \frac{A}{A_c}.
\label{eq:kappa_neel_clean}
\eeq
For $A \rightarrow 0$, \ie\ vanishing antiferromagnetic order, we recover the 
result by Durst and Lee~\cite{DurstLee_PRB2000} as expected.

Tuning the order parameter $A$ at fixed chemical potential beyond a critical 
value $A_c$, the nodes can collide and become gapped as discussed in 
Appendix~\ref{nodalCol}. This entails an exponential suppression of the heat 
conductivity due to the resulting gap in the spectrum in the absence of large 
disorder broadening. This scenario could be relevant in the strongly 
underdoped regime, as gapping out the nodes leads to a phase transition from a 
superconductor to a half-filled insulator. 

The above result also indicates that close to the doping $p^\ast$ where 
antiferromagnetic order appears, the number of nodes and the nodal velocities 
remain unaffected across the phase transition. Thus, a smooth behavior of the 
heat conductivity is expected near $p^\ast$, consistent with our numerical 
results in Fig.~\ref{fig:CleanDirtyLimits_dSCAF}.  

A few further comments are in order. The apparent divergence of $\kappa / T$ 
for $A \rightarrow A_c$ is an artifact of the clean approximation, which will 
get smoothened out by disorder. Moreover, there is no nematic order 
(corresponding to the breaking of the $C_4$ symmetry of the square lattice to 
$C_2$) and $\kappa_{xx} = \kappa_{yy}$. This is markedly different from the 
cases of superconductivity coexisting with spiral antiferromagnetism or charge 
density waves with axial wave-vector~\cite{DurstSachdev_PRB2009}. Finally, 
our result is valid for any anisotropy ratio $v_F/v_{\Delta}$, unlike the 
isotropic limit discussed in Ref.~\onlinecite{DurstSachdev_PRB2009}. The 
dependence of $\kappa$ on the order parameter magnitude is also different in these two 
cases, and may be used as a probe to distinguish between these two different 
orders in a clean $d$-wave superconductor. 

Note that although the results in Ref.~\onlinecite{DurstSachdev_PRB2009} correspond to a s-wave charge density wave, the wave-vector $(\pi,0)$ is obtained considering the second harmonic of the experimentally observed wave-vector of $(\pi/2,0)$. For a $d$-wave bond density wave which has been observed in STM experiments \cite{hamidian_2015nphys}, we need to consider the second harmonic which has a squared form factor, i.e, $\psi_{\k} \sim (\text{cos}k_x - \text{cos}k_y)^2$. Therefore, the equation $\partial_{\k}\psi_{\k} = 0$ still holds at the nodes which lie along $k_x = \pm k_y$, and their results are valid for the d-form factor density wave state as well. 

\subsubsection{Spiral order}

For the case of spiral order, the metallic state with no superconductivity has 
only two hole pockets, which are in the region $k_x > 0$ for $\Q = (\pi - 2\pi 
\eta, \pi)$. This can be understood from the fact that for small $A$, the 
particle-hole polarization bubble at momentum $\Q$ is maximum when $\Q$ 
approximately nests two segments of the Fermi surface, and accordingly the 
saddle-point free energy for the fluctuations of the order-parameter field after integrating out the 
fermions is minimum. As we fix the phenomenological doping dependence of the 
order parameter amplitude $A$ and minimize the free energy by optimizing 
the incommensurability $\eta$~\cite{Eberlein2016}, the preferred $\eta$ gaps out a large 
part of the Fermi pockets to reduce the mean-field free energy. 

Adding superconductivity on top of the spiral state therefore implies that there 
are only four nodes in the extended BZ, coming from the two hole-pockets. The 
other four nodes will collide and disappear once spiral order sets in. 
Thus, in the clean limit we expect the thermal conductivity to drop to half 
of its original value soon after crossing $p^\ast$ from the overdoped side. 
Evaluating the heat conductivity by focusing on the vicinity of the four 
nodal points for $p \lesssim p^*$, the thermal 
conductivity for the clean $d$-wave superconductor with spiral order is given by 
half of the Durst-Lee value,
\beq
\frac{\kappa_{ii}(\Omega \rightarrow 0,T)}{T} =  \frac{k_B^2 (v_F^2 + v_{\Delta}^2)}{6 v_F v_{\Delta}} 
\eeq
Indeed, numerical results in Fig.~\ref{fig:CleanDirtyLimits_dSCAF} show a sharp drop of the thermal conductivity by a factor of two across the critical point. 

\subsection{Violation of the Wiedemann-Franz law}

In the clean limit, we can evaluate the bare-bubble electrical conductivity due to gapless nodal quasiparticles as 
described in section \ref{afWF}. For N\'{e}el order, the non-superfluid contribution to the electrical current is 
given by:
\beq
 \j^{e}(\q \rightarrow 0, \Omega) =  \sum_{\k ,\omega}^{\prime} \Psi^{\dagger}_{k} \mathbf{V}_{\k} \psi_{k + q}, \text{ where } \nn
\mathbf{V}_{\k} = \begin{pmatrix}
 \vf(\k) \tau_{0} & 0 \\
 0 &  \vf(\k + \Q)  \tau_{0}
 \end{pmatrix} \approx \begin{pmatrix}
 \vf(\k)  \tau_{0} & 0 \\
 0 &  -\vf(\k)  \tau_{0}
 \end{pmatrix}
\eeq
From this, the quasiparticle contribution to the electrical conductivity (denoted by $\tilde{\sigma}$) can be evaluated using an analogous computation to the thermal conductivity:
\beq
\tilde{\sigma}_{ii} (\Omega \rightarrow 0,T \rightarrow 0) &=& \frac{e^2}{\pi^2} \int \frac{d^2k}{8\pi} \text{Tr}\left[ G_R^{\prime \prime}(\k, 0)\mathbf{V}_{\k} G_R^{\prime \prime}(\k, 0)\mathbf{V}_{\k}\right] \nn
& = & \frac{e^2}{\pi^2} \frac{v_F}{v_{\Delta}}  \sqrt{1 - \alpha^2} \,  \Theta(1 - \alpha), \text{ where } \alpha = \frac{A}{A_c}
\eeq
Therefore, in the clean limit we have:
\beq
\frac{\kappa}{\tilde{\sigma} T} = \frac{\pi^2 k_B^2}{3 e^2} \left( 1 + \frac{1}{1 - \alpha^2} \frac{v_{\Delta}^2}{v_F^2}\right)
\eeq
Since the Fermi surface is modified as a function of doping, $\alpha$, the Fermi velocity and the gap velocity all change  and therefore $\kappa/\tilde{\sigma} T$ is not a constant as a function of doping in the antiferromagnetic ($A \neq 0$) regime. However, typically the Fermi velocity is much larger than the gap velocity, and therefore this correction is expected to be small. As $\alpha \rightarrow 1$, the correction appears large if we hold $v_{\Delta}$ fixed. But this is a rather unphysical limit as increasing antiferromagnetism to its critical value will reduce the superconductivity $\Delta_d$, and $v_{\Delta}$ will also drop. Note that for isotropic disorder scattering, the single-particle lifetime is equal to the scattering time for free fermions. Even for our lattice model, we expect only minor modifications from the bare-bubble result due to vertex corrections. In particular, in the dirty limit $v_{\Delta}/v_F \rightarrow 0$ and the Wiedemann-Franz law is exactly satisfied by the quasiparticle contribution to the electric current, as long as the disorder is relatively weak compared to the Fermi energy, as described in section \ref{afWF}.

\subsection{Numerical results}
\begin{figure}
	\centering
	\includegraphics[width=0.5\linewidth]{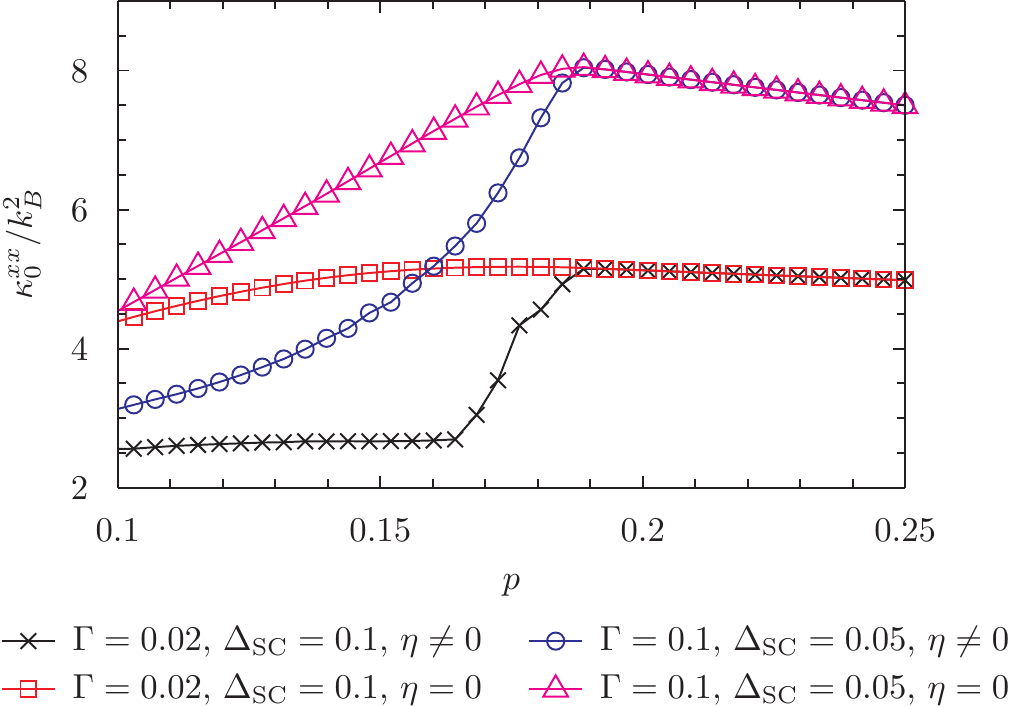}
	\caption{Doping evolution of the heat conductivity in the clean and dirty 
limits, comparing commensurate and incommensurate antiferromagnetic order. 
Parameters are $t' = -0.35$ and $A_{p_\text{min}} = 1.0$.}
	\label{fig:CleanDirtyLimits_dSCAF}
\end{figure}
\begin{figure}
	\centering
\includegraphics[width=0.5\linewidth]{%
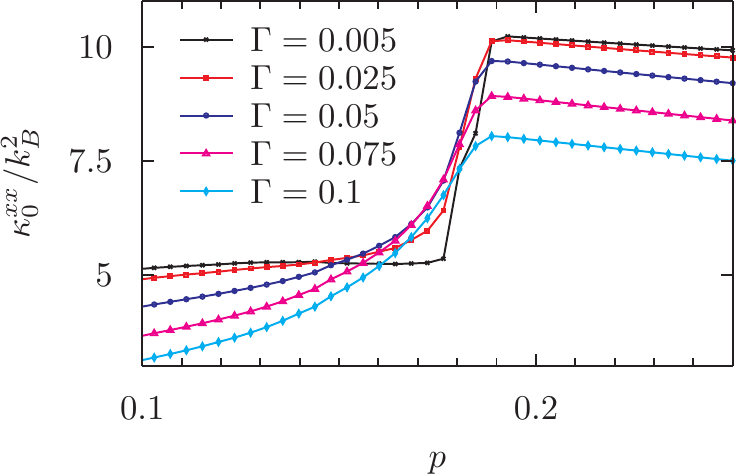}
	\caption{Evolution of heat conductivity from the clean to the dirty limit for 
coexisting superconductivity and incommensurate antiferromagnetism, for 
$A_{p_\text{min}} = 1.0$ and $\Delta = 0.05$.}
	\label{fig:Gamma_evolution}
\end{figure}

In Fig.~\ref{fig:CleanDirtyLimits_dSCAF}, we compare the doping dependence of 
the heat conductivity in the clean and dirty limit for antiferromagnetism 
coexisting with superconductivity. In the case of commensurate 
antiferromagnetism, the change near $p^\ast$ is less pronounced than in the 
case of incommensurate antiferromagnetism for both the clean and dirty limits. 
In the clean limit, the location of $p^\ast$ is not discernible from 
the plot of $\kappa_0^{xx}$ in the commensurate case. This is consistent with 
the analytical result in 
Eq.~\eqref{eq:kappa_neel_clean}. In contrast, at the doping where the spiral 
antiferromagnetic order appears, the heat conductivity drops to roughly half of 
its value. In the case of incommensurate antiferromagnetism coexisting with 
superconductivity in the dirty limit, the doping dependence of the heat 
conductivity is much smoother, as already discussed above. In 
Fig.~\ref{fig:Gamma_evolution}, we show the evolution of the heat conductivity 
for various scattering rates from the clean to the dirty limit, demonstrating 
how the jump gets washed out with increasing scattering rate.

\section{Influence of doping-dependent scattering in the dirty limit}
\label{sec:DopingDependentScattering}
In Sec.~\ref{sec:AnalyticAFMetal} we discussed that the drop in the thermal 
conductivity across the antiferromagnetic phase transition can be understood 
in terms of a reconstruction of the Fermi surface in case of a disordered 
antiferromagnetic metal or a dirty superconductor ($\Delta_d \ll 
\Gamma$). In this scenario, the relative drop across the phase transition is 
smaller than the drop of the Hall number, whereas experiments suggest that the 
drops are of very similar magnitude. The experiments have been interpreted in terms of a Drude-like 
model, which allows to connect the drop in the conductivity with a drop in the 
charge-carrier density~\cite{Collignon2016} by assuming that the charge 
carrier mobility is unchanged across the phase transition. However, this 
requires the validity of an effective mass picture, or nearly circular 
pockets, which only holds for a very large antiferromagnetic order parameter. 
Below the optimal doping QCP, the pockets are quite distorted and an effective mass 
picture is therefore not appropriate. Very recent experiments on electron-doped cuprate LCCO \cite{Sarkar2017-arXiv}
have also observed similar resitivity upturns which cannot be explained only by a drop in carrier density. 
In this section, we provide an alternative scenario that would explain the larger drop in conductivity.

The key observation is that the scattering rate $\Gamma$ cancels in the 
Hall number~\cite{Voruganti1992,Eberlein2016}, while the thermal conductivity 
in the dirty superconductor is approximately proportional to $\Gamma^{-1}$. 
Therefore, one might anticipate that additional sources of scattering that 
appear once antiferromagnetism sets in can entail a larger drop of the thermal 
conductivity. We show that a phenomenological doping-dependent scattering rate 
can indeed lead to similar behavior and drop sizes in the Hall number and the 
conductivities. We 
then argue for a possible source of enhanced scattering in the underdoped 
regime.

\begin{figure}
	\centering
	\includegraphics[width=0.5\linewidth]{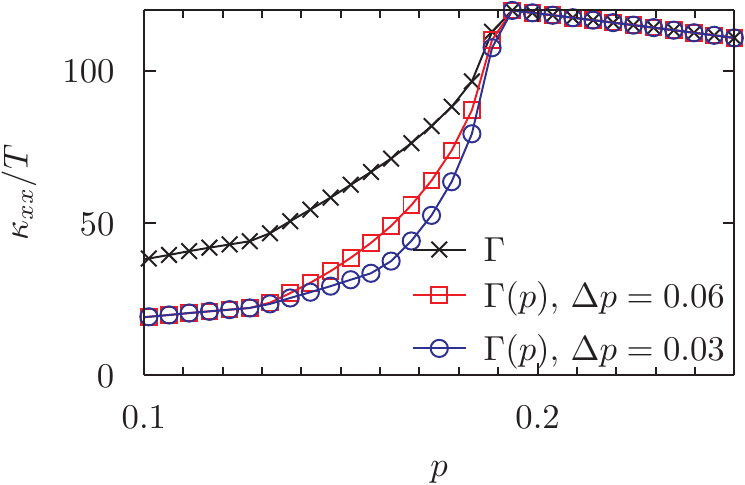}
	\caption{Doping dependence of the heat conductivity for $t' = -0.35$, 
$A_{p_\text{min}} = 1.0$ and $\Gamma_0 = 0.01$. The curve labeled $\Gamma$ 
shows the same data as in Fig.~\ref{fig:Comparison_iAF_cAF} for comparison. The 
other two curves were obtained for a doping dependent scattering rate that 
doubles over a doping range of $\Delta p$ below $p^\ast$.}
	\label{fig:kappa_Gamma_p}
\end{figure}
For simplicity, we assume that the doping dependence of $\Gamma(p)$ is given by
\begin{equation}
	\Gamma(p) = \begin{cases}
	            	2 \Gamma_0 	&	\text{for } p < p^\ast - \Delta p\\
								2 \Gamma_0 - \Gamma_0 (p - p^\ast + \Delta p) / \Delta p
& \text{for } p^\ast - \Delta p \leq p \leq p^\ast \\
								\Gamma_0		&	\text{for } p > p^\ast,
	            \end{cases}
\end{equation}
where $\Delta p$ is the doping range over which the scattering rate increases. 
\begin{figure}
	\centering
	\includegraphics[width=0.5\linewidth]{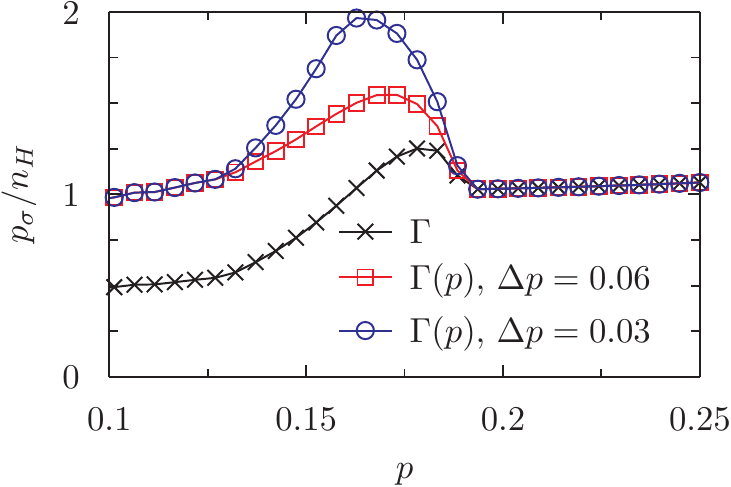}
	\caption{Ratio of the charge carrier concentrations as extracted from 
the conductivity ($p_\sigma$) and the Hall coefficient ($n_H$) as a function of 
doping. We chose $A_{p_\text{min}} = 1.0$, where electron pockets exist for 
$0.13 \leq p \leq 0.19$. $\Delta p$ is the doping range over which $\Gamma(p)$ 
doubles with decreasing doping.}
	\label{fig:pSigma_nH_Gamma}
\end{figure}
In Fig.~\ref{fig:kappa_Gamma_p} we show results for the heat conductivity of a 
disordered antiferromagnetic metal as obtained with this choice of $\Gamma(p)$. 
We mentioned in the 
last sections that for $\Gamma(p) = \Gamma_0$ discrepancies between experiments 
and the theory of transport in a
superconductor in the dirty limit showed up in the Hall angle and the doping 
dependence of the charge carrier density. In Fig.~\ref{fig:pSigma_nH_Gamma}, we 
show the ratio between the charge carrier density as extracted from the 
conductivity and the Hall number. For $\Gamma(p) = \Gamma_0$, with decreasing 
doping we find a small peak and then a decrease to values significantly below 
one. Adding doping dependent scattering, the peak at $p < p^\ast$ increases as 
the conductivity drops faster. Note that in this section we assume that $R_H$ 
depends only weakly on the scattering rate for $\Gamma_0 \ll E_F$, where $E_F$ 
is the Fermi energy. Our results for $p_\sigma / n_H$ can be compared with the 
experimental results by Collignon~\etal~\cite{Collignon2016}, yielding 
good qualitative agreement.

\begin{figure}
	\centering
	\includegraphics[width=0.5\linewidth]{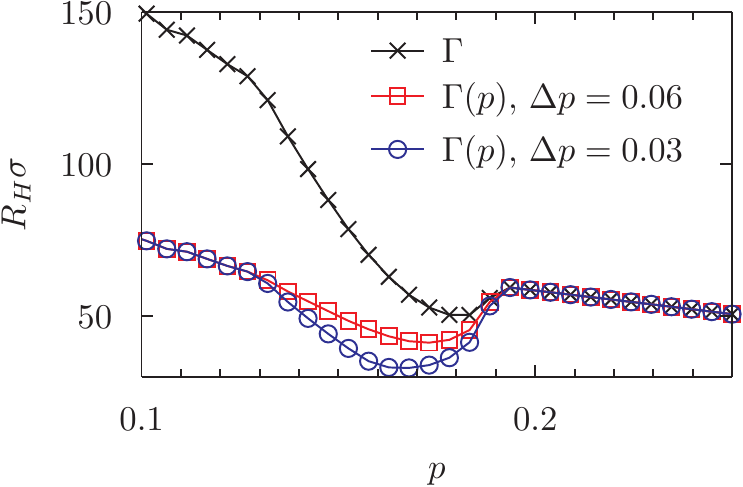}
	\caption{Hall angle as a function of doping computed for different 
doping-dependences of the scattering rate. We chose $A_{p_\text{min}} = 1.0$, 
where electron pockets exist for 
$0.13 \leq p \leq 0.19$. $\Delta p$ is the doping range over which $\Gamma(p)$ 
doubles with decreasing doping.}
	\label{fig:RH_sigma_Gamma_p} 
\end{figure}

A similar picture emerges from the Hall angle $R_H \sigma$. In 
$\textrm{Nd-LSCO}$, a drop by a factor of three is observed over the width of 
the transition with decreasing doping~\cite{Collignon2016}. In the 
disordered antiferromagnetic metal, a rather small drop is observed in this 
quantity for $p < p^\ast$, followed by an increase at smaller $p$. As can be 
seen in Fig.~\ref{fig:RH_sigma_Gamma_p}, adding doping dependent scattering 
allows to enhance the size of the drop and weakens the decrease at smaller 
doping, leading to a better qualitative agreement with the experimental 
results~\cite{Collignon2016}.

A doping-dependent scattering rate that increases for $p < p^\ast$ can thus 
improve the qualitative agreement between theory and experiment in various 
transport properties. In the following we argue that the competing ordering 
tendencies at different energy scales in underdoped cuprates could provide a 
mechanism for such a doping-dependence of the scattering rate. In La-based 
cuprates like Nd-LSCO, at low dopings ($p \sim 0.12$) there is evidence of 
stripe-like ordering from neutron scattering and X-ray spectroscopy 
\cite{Tranquada_Nature95,Tranquada_PRB96,Comin16}. In other cuprate materials 
like BSCCO, local patches of charge modulations have been seen in STM 
experiments \cite{Fujita_PNAS2014,FujitaScience2014,hamidian_2015nphys}. 
Theoretical studies also show that the reconstructed small Fermi surface has a 
dominant instability towards bond-density waves (BDW) at an incommensurate 
wave-vector
with a $d$-wave form factor \cite{DCSS14,SSDC16}. The results of recent transport 
experiments~\cite{Laliberte2016-arXiv,Collignon2016} suggest that 
charge-ordering sets in at a lower doping than $p^\ast$, where the pseudogap 
line terminates. However, short-range charge density modulations seem to 
be omnipresent in the pseudogap phase, and could act as additional sources of 
scattering. Indeed, time-reversal symmetric disorder can destroy long range 
density wave order in the charge channel as it couples linearly to the order 
parameter, but it can only couple quadratically to the spin 
density wave order parameter and is therefore a less relevant perturbation to 
antiferromagnetic order~\cite{LNAMSK_2017}. Below, we explore a simple model of 
disordered density waves and estimate its contribution to the quasiparticle 
scattering rate.

\begin{figure}[h!]
	\centering
	\includegraphics[width=5cm,height=4.5cm]{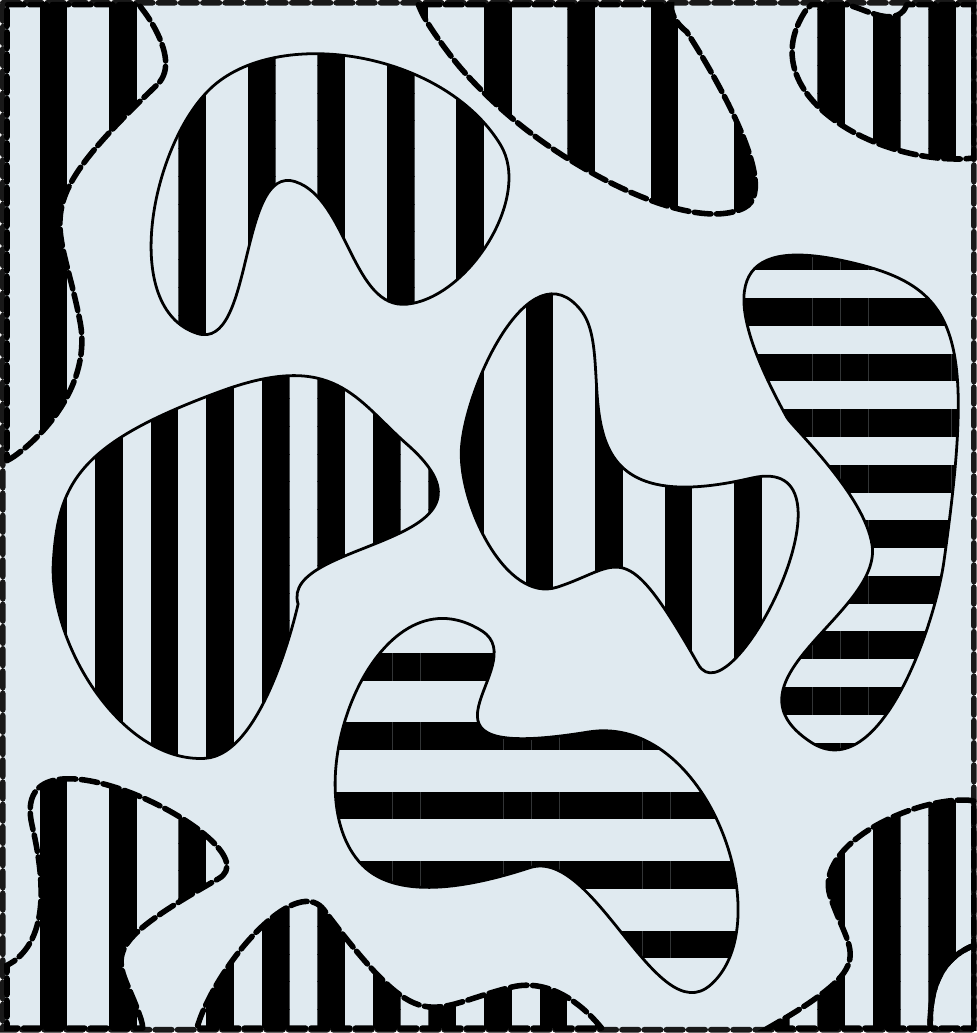}
	\caption{Patches of density waves that act as additional quenched disorder.}
	\label{fig:DW}
\end{figure}
We model the disorder-induced scattering as arising from pinned short-range 
charge-density wave order with a domain size given by the correlation length 
$\xi$, which we assume to be of the order of ten lattice spacings. Each domain 
is locally unidirectional with an incommensurate ordering 
wave vectors $\Q$, as shown in Fig.~\ref{fig:DW}. The order parameter 
$P_{ij}$ scatters an 
electron from momentum $\k - \Q/2$ to $\k + \Q/2$. Therefore, each such patch 
can be considered to be a short-range potential scatterer with the appropriate 
matrix element for scattering between electron states $\ket{\k}$ and 
$\ket{\k^{\prime}}$ being given by $P_{\Q} \, f\left[(\k + \k^{\prime})/2\right]$, where 
$f(\k) $ is some appropriate internal form factor. We assume a 
phenomenological Lorentzian dependence on $\Q$ that it peaked at $\Q_0$, the 
ordering wave-vector with the largest susceptibility. Assuming weak disorder 
with a density $n_i \sim 0.01$, we can self-average over the disorder to find 
a scattering time $\tau_{2}$ given in the Born approximation by (assuming it is 
independent of initial state)
\beq
\tau_2^{-1} &=& 2 \pi  \int \frac{d^{2}\Q}{4 \pi^2} g(\Q)  \int \frac{d^{2}\k}{4 \pi^2} \big| P_{\Q}(\k, \k^{\prime}) f((\k + \k^{\prime})/2) \big|^2 \delta(\xi_{\k} - \xi_{\k^{\prime}}), \text{ where } \nn
P_{\Q}(\k, \k^{\prime}) &\sim& \frac{1}{\xi^{-2} + (\k - \k^{\prime} - \Q)^2},
\eeq 
where $g(\Q)$ is a normalized function peaked at $\Q_0$. Now a gradual increase 
in the density wave correlation length $\xi$ with decreasing doping can result 
in a larger scattering rate. Saturation of the charge-density wave correlation 
length due to disorder also entails saturation of the scattering rate.


\section{Topological order in the pseudogap phase}
\label{sec:topo}

\begin{figure}
\begin{center}
\includegraphics[height=4in]{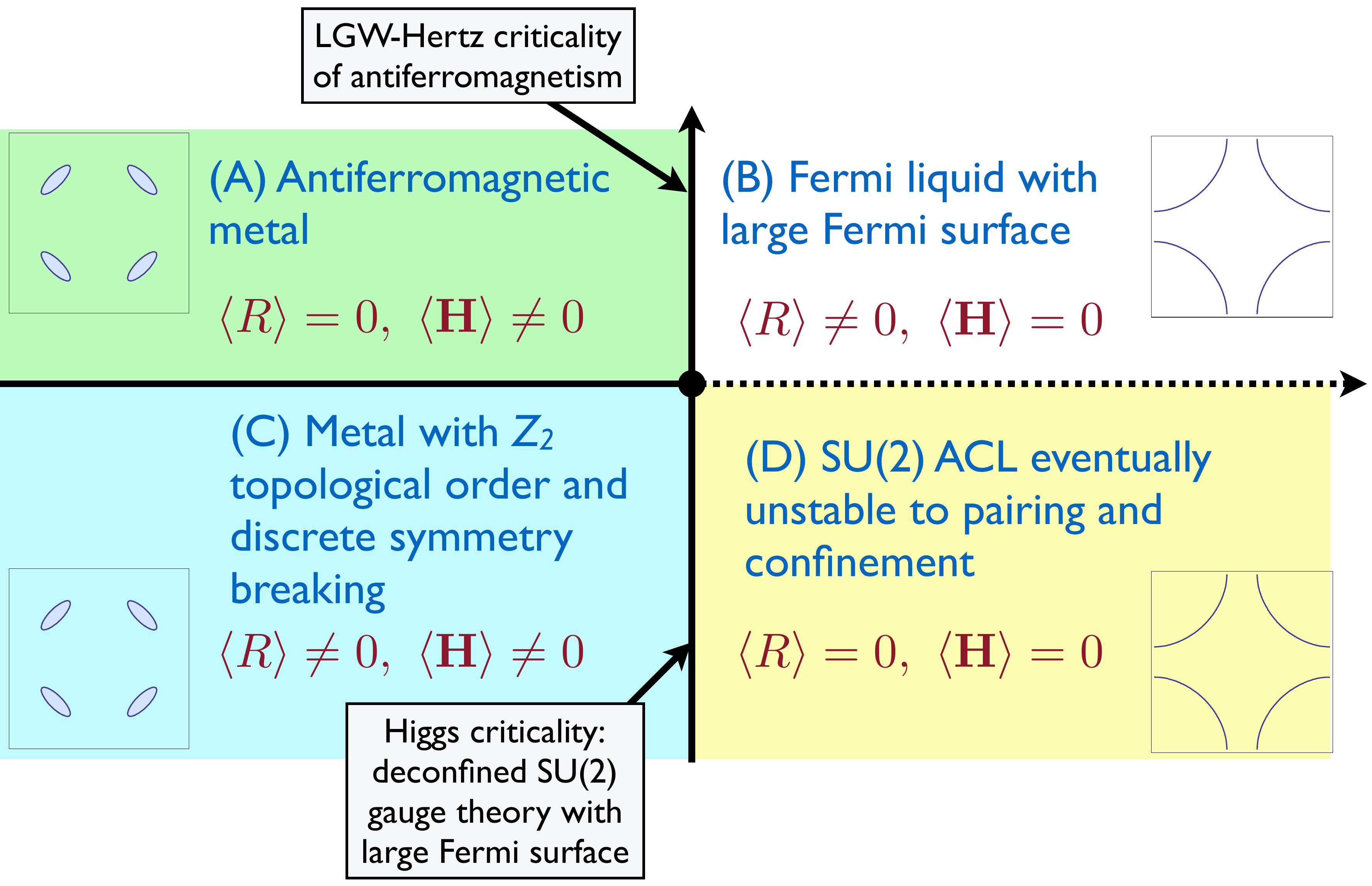}
\end{center}
\caption{Phase diagram adapted from Refs.~\onlinecite{SSDC16,SCSS17}. The $x$ and $y$ axes are parameters controlling the condensates of 
$\mathbf{H}$ and $R$ respectively. There is long-range antiferromagnetic order only in phase A, where both $R$ and 
$\mathbf{H}$ condensates are present. Phase C is a model for the pseudogap with topological order. We argue in the text that,
in the simplest theory, the charge and energy transport properties of the A $\rightarrow$ B transition are identical
to those of the C $\rightarrow$ D transition. The dashed line between phases C and D represents a crossover.}
\label{fig:phasediag}
\end{figure}
In this paper, we have so far discussed electrical and thermal transport in a 
mean-field model for a quantum phase transition from a (in-) commensurate antiferromagnet to a 
non-magnetic Fermi liquid. In Fig.~\ref{fig:phasediag}, 
we present the phase diagram of a SU(2) lattice gauge theory \cite{SSDC16,SCSS17} of the square lattice
Hubbard model, in which such a quantum transition
corresponds to taking the route A $\rightarrow$ B with increasing doping. Along this route,
the optimal doping criticality is associated with the Landau-Ginzburg-Wilson-Hertz \cite{hertz} theory of
the antiferromagnetic quantum critical point.
However, there have, so far, been no indications that long-range
antiferromagnetic order is present in the pseudogap regime of the hole-doped cuprates.
So we instead examine the route A $\rightarrow$ C $\rightarrow$ D $\rightarrow$ B in Fig.~\ref{fig:phasediag} as describing the
evolution of phases with increasing hole-doping. In this route, the pseudogap is phase C, a metal with $\mathbb{Z}_2$ topological
order, and the optical doping criticality is the topological phase transition between phases C and D.

The specific scenario illustrated in Fig.~\ref{fig:phasediag} assumes that the pseudogap is a 
$\mathbb{Z}_2$ algebraic 
charge liquid ($\mathbb{Z}_2$-ACL) or a $\mathbb{Z}_2$ fractionalized Fermi 
liquid ($\mathbb{Z}_2$-FL$^\ast$) \cite{SSDC16,SBCS_2016}. For both these 
phases, the only low energy quasiparticles are charge-carrying fermions with a 
small Fermi surface. These phases can be described as metals with quantum-fluctuating
antiferromagnetism in the following manner. We introduce a spacetime-dependent SU(2) spin rotation, 
$R_i$ to transform the electron operators $c_{i \alpha}$ into `rotated' fermions $\psi_{is}$, with $s=\pm$:
\beq
\left( \begin{array}{c} c_{i\uparrow} \\ c_{i\downarrow} \end{array} \right) = R_i \left( \begin{array}{c} \psi_{i,+} \\ \psi_{i,-} \end{array} \right),
\label{R}
\eeq
where 
\beq
R_i^\dagger R_i = R_i R_i^\dagger = 1. 
\eeq
The same transformation rotates the local magnetization $\mathbf{m}_i$ to a `Higgs' field $\mathbf{H}_i$
\beq
{\bm \sigma} \cdot \mathbf{m}_i = R_i \, ({\bm \sigma} \cdot \mathbf{H}_i )\, R_i^\dagger \label{mR}
\eeq
Note that under Eq.~(\ref{mR}), the coupling of the magnetic moment $\mathbf{m}_i$ to the electrons
is equal to the coupling of the Higgs field to the $\psi$ fermions
\beq
\mathbf{m}_i \cdot c_{i \alpha}^\dagger {\bm \sigma}_{\alpha\beta} c_{i \beta}^{\vphantom \dagger} = 
\mathbf{H}_i \cdot \psi_{i s}^\dagger {\bm \sigma}_{ss'} \psi_{i s'}^{\vphantom \dagger} \label{mH}
\eeq
A second key observation is that the mappings in Eqs.~(\ref{R}) and (\ref{mR}) are invariant under the SU(2) gauge transformation
generated by $V_i$, where
\begin{eqnarray}
\left( \begin{array}{c} \psi_{i,+} \\ \psi_{i,-} \end{array} \right) &\rightarrow& V_i (\tau) \left( \begin{array}{c} \psi_{i,+} \\ \psi_{i,-} \end{array} \right) \nn
\quad R_i &\rightarrow R_i& V_i^\dagger (\tau) \nn
{\bm \sigma} \cdot \mathbf{H}_i  & \rightarrow &   V_i \,( {\bm \sigma} \cdot \mathbf{H}_i )\, V_i^\dagger  . \label{gauge}
\label{gauge2}
\end{eqnarray}
So the resulting theory for the $\psi$, $R$ and $\mathbf{H}$ will be a SU(2) lattice gauge theory.

We are interested here in the properties of state C as a model for the pseudogap. From Fig.~\ref{fig:phasediag}, we observe
that in this state the local antiferromagnetic order $\mathbf{m}_i$ quantum fluctuating, but the Higgs field
(which is the antiferromagnetic order in a rotating reference frame) is a constant. Moreover, from Eq.~(\ref{mH}), the coupling
of the $\psi$ fermions to the Higgs field is identical to the coupling of the electrons to the physical magnetic moment. 
If we assume that the dispersions of the $\psi$ and $c$ fermions are the same (in a suitable gauge), then we can compute
the charge and energy transport properties of the transition from state C to state D without further analysis: they are identical to
the charge and energy transport properties of the transition from state A to state B which were computed in earlier sections
of this paper. There are significant differences in the spin transport properties of C $\rightarrow$ D from A $\rightarrow$ B,
but these have not so far been experimentally accessible in the cuprates.

So we have the important conclusion that the concurrence between theory and experiments, in this paper and in earlier
work \cite{Storey2016,Eberlein2016}, applies also for the 
topological phase transition, C $\rightarrow$ D, model of the optical doping criticality. And this model has the important 
advantage that long-range antiferromagnetic order is not required in the pseudogap phase C. Phase D is described by
a SU(2) gauge field coupled to a large Fermi surface of fermions with SU(2) gauge charges: such a phase is expected
to be unstable to a superconductor in which all SU(2) gauge charges are confined, and so the state is formally the same
as a BCS superconductor. However, a magnetic field could suppress the superconductivity and expose the underlying non-Fermi liquid,
and this makes phase D a candidate to explain the observed strange metal in the overdoped regime \cite{Cooper603}.

We note that it is also possible to construct models of 
$\mathbb{Z}_2$-FL$^\ast$, different from that in Fig.~\ref{fig:phasediag}, 
building on the models reviewed in Ref.~\onlinecite{PALRMP} in which the low-energy charge-carrying excitations are 
bosonic~\cite{Punk15,SCYQSSJS_2016}. However, these models support 
charge-neutral spinons in the deconfined phase~\cite{Chatterjee2016} and 
therefore violate the Wiedemann-Franz law 
quite strongly. The reason is that the spinons contribute to the 
thermal conductivity but not to the electrical conductivity. Such violations 
have not been observed in experiments~\cite{Grissonnanche2016}.

\section{Discussion}
\label{sec:dis}
It is natural to ask how the presence of other excitations or 
fluctuations would affect our findings above. The parameter regime very close to the critical 
doping $p^\ast$ cannot be reliably described by our simple mean-field approach. 
This requires a more sophisticated theory of transport in a strange metal and 
the consideration of quantum critical fluctuations. However, the quantum critical regime of doping shrinks to a point 
as $T \rightarrow 0$, and away from this regime fermionic quasiparticles exist and are well-described 
by nearly non-interacting fermions. Interaction effects can be taken into account by Fermi liquid corrections \cite{DurstLee_PRB2000} to the conductivities. In principle, vertex corrections and Fermi liquid corrections may be different on the two sides of the phase transition. We argued that the results for the conductivities in 
the dirty limit can be well understood under the assumption that the Fermi 
velocity does not change across the optimal doping QCP. There is experimental evidence for certain cuprates like BSSCO that the Fermi velocity is roughly constant across $p^*$, although it does get renormalized to smaller values for lower doping \cite{Sebastian2011,Vishik_ARPESreview}. Our results are thus robust if the Fermi velocity in the calculation is interpreted as the measured Fermi velocity from experiments. Fermi liquid corrections should thus not change our conclusions substantially. Moreover, we studied the interaction of fermions with disorder only within a simple relaxation time approximation. We expect this to capture the qualitative features in the relevant limits. It would be 
interesting to determine the scattering time self-consistently, as gapping out 
parts of the Fermi surface may also influence the scattering time. This could 
be done in an unconstrained Hartree-Fock calculation similar to Ref.~\onlinecite{Hirschfeld2009}.

On the overdoped side, there are no gapless excitations besides the fermions. In 
the scenario where static iAF order disappears at the QCP, in the overdoped 
regime $p > p^\ast$ magnons are gapped and do not contribute to the heat 
conductivity at low temperatures. On the overdoped side of the QCP from the 
topological metal to the normal metal, there are also no additional low energy 
excitations which could contribute to thermal transport. The reason is that in 
this scenario, the normal metal is a confined phase, where all additional excitations carrying 
$\mathbb{Z}_2$ gauge charge are confined and the gauge field is massive.

Similar arguments hold in the underdoped regime. In the $\mathbb{Z}_2$-ACL or 
$\mathbb{Z}_2$-FL$^\ast$ with fermionic chargons, the charge-neutral 
spin-excitations as well as the visons (which are the $\mathbb{Z}_2$ gauge 
fluxes) yield additional contributions to the heat conductivity at finite 
temperature. However, these are suppressed at low temperatures because the 
spinons and visons are both gapped. In the scenario with static 
iAF order, magnons yield additional contributions to the heat conductivity, 
which vanish at zero temperature. The two scenarios could possibly be 
distinguished at finite temperature, where gapped gauge fields contribute 
differently from magnons. Note that no magnon contribution to the heat 
conductivity of $\mathrm{YBCO}$ is seen beyond the doping where 
long-range commensurate antiferromagnetic order at finite temperature 
disappears in the strongly underdoped regime~\cite{Doiron-Leyraud2006}. This 
could, however, also be a consequence of long-range incommensurate 
antiferromagnetic order only existing in the ground state~\cite{Haug2010}. We 
leave the study of this interesting problem for future work.

\section{Conclusions}
\label{sec:conc}

We summarize the main findings of our numerical computations for the electrical and thermal conductivities, and their
relationship to observations. 

In Fig.~\ref{fig:Influence_iAF}, we showed the doping dependence of the thermal conductivity
of metallic states in the presence
of spiral antiferromagnetic order at low doping. The comparison of these results with the commensurate antiferromagneticsm case appears
in Fig.~\ref{fig:Comparison_iAF_cAF}. Although there is a difference between these cases, both sets of results show
that the drop in the thermal conductivity $\kappa$ between large and small $p$ is smaller than that found for the Hall effect in Ref.~\onlinecite{Eberlein2016},
as shown in Fig.~\ref{fig:Comparison_pSigma_nH}. These results are at odds with the recent observations of Collignon~\etal\
\cite{Collignon2016,Michon2017} who found the same drop in the carrier density in the thermal conductivity and the Hall effect.

Next we turned to corresponding computations in the presence of superconductivity. In Fig.~\ref{fig:CleanDirtyLimits_dSCAF}, we plotted the evolution of thermal conductivity as a function of doping in the clean and dirty limits. In the clean limit, $\kappa$ shows markedly different behavior for commensurate and incommensurate antiferromagnetic order. The appearance of N\'{e}el order entails a gradual drop of $\kappa$ on the underdoped side ($p < p^*$), whereas advent of incommensurate spiral order results in a sharp drop by a factor of two, consistent with our analytical results. Figure~\ref{fig:Gamma_evolution} depicts how this sharp drop is smoothened out as a function of increasing disorder. We also noted that appearance of antiferromagnetism can be distinguished from other orders (like charge density wave \cite{DurstSachdev_PRB2009}) co-existing with superconductivity by studying the evolution of the thermal conductivity across the quantum critical point.  

We then discussed how a doping-dependent scattering rate, possibly due to quenched density fluctuations, affects the thermal conductivity in the disordered antiferromagnetic metal or the dirty superconductor. Figure~\ref{fig:kappa_Gamma_p} shows the evolution of $\kappa$ for different doping dependent scattering rates. A comparison of the carrier densities extracted from conductivity and the Hall effect appears in  Fig.~\ref{fig:pSigma_nH_Gamma}, and a plot of the Hall angle as a function of doping is shown in Fig.~\ref{fig:RH_sigma_Gamma_p}; both are in good qualitative agreement with recent experimental data of Collignon~\etal\ and Michon \etal\ \cite{Collignon2016,Michon2017}.

Finally, we presented an alternate description of the pseudogap phase as an exotic metal with $\mathbb{Z}_2$ topological order, but without long range antiferromagnetism. Figure~\ref{fig:phasediag} shows a phase diagram outlining the two distinct routes from a small Fermi surface in the pseudogap phase to a large Fermi surface on the overdoped side. We argued that in both cases, the electrical, Hall \cite{Storey2016,Eberlein2016} and thermal conductivities exhibit identical evolution as a function of doping at low temperatures, and therefore the observations in Refs.~\onlinecite{Collignon2016,Michon2017} can be equally well explained by a phase transition from a regular Fermi liquid to a topological pseudogap phase.

\section*{Acknowledgments} 
We would like to thank L.~Taillefer for valuable discussions. SC was supported 
by the Harvard-GSAS Merit Fellowship. SS acknowledges support from Cenovus 
Energy at Perimeter Institute. AE acknowledges support from the German National 
Academy of Sciences Leopoldina through grant LPDS~2014-13. Moreover, AE would 
like to thank the Erwin Schr\"odinger International Institute for Mathematics and 
Physics in Vienna, Austria, for hospitality and financial support during the 
workshop on ``Synergies between Mathematical and Computational Approaches to 
Quantum Many-Body Physics''. This research was supported by the NSF under Grant 
DMR-1360789 and MURI grant W911NF-14-1-0003 from ARO. 
Research at Perimeter Institute is supported by the Government of Canada through 
Industry Canada and by the Province of Ontario through the Ministry of Economic 
Development \& Innovation.

\appendix

\section{N\'{e}el ordered $d$-wave superconductor and the phenomenon of nodal 
collision}
\label{nodalCol}
\begin{figure}[htbp!]
\includegraphics[scale=0.8]{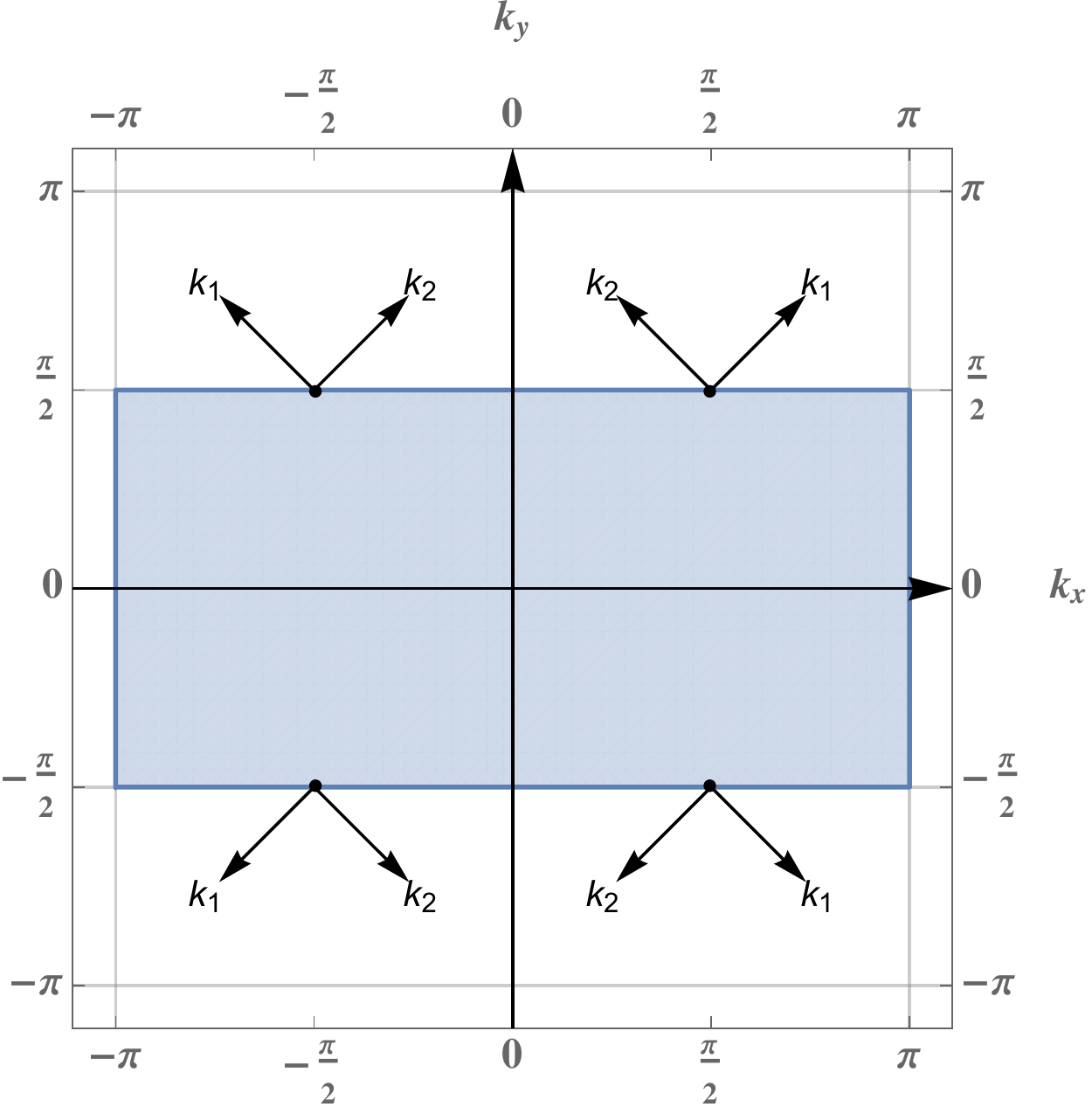}
\caption{Local coordinate systems adapted to the nodes, defined about the four nodal collision point $(\pm \pi/2, \pm \pi/2)$. The blue portion is the reduced BZ used in our calculations.}
\label{coord}
\end{figure}

We take a more careful look at the nodes of the general dispersion described in 
Eq.~(\ref{cafEigs}) for $\Q = (\pi,\pi)$, with a bare fermionic dispersion 
$\xi_{\k}$ that pertains to the band structure of the overdoped cuprates. To 
model this, we choose the nodes to lie along the diagonal likes $k_x = \pm k_y$
at a distance $k_F$ from the origin. We assume that the node $\K_0$ in the first quadrant of the BZ 
is separated from the $(\pi/2,\pi/2)$ point by a small distance $k_o$, following 
Ref.~\onlinecite{DurstSachdev_PRB2009}. The precise criteria is that the node lies quite
close the center of the positive quadrant of the BZ, and the distance 
$k_o = \pi/\sqrt{2} - k_F$ is much smaller than $k_F$. Within this approximation, we see the 
same phenomenon of nodal collision which gaps out the nodal quasiparticles of the $d$-wave 
superconductors, as described for charge density waves in 
Ref.~\onlinecite{DurstSachdev_PRB2009}. The diagonally opposite nodes collide at 
the boundary of the reduced BZ beyond a certain critical value of the order 
parameter $A$ (which we determine below), and this renders the spectrum fully 
gapped. Since we are mainly interested in the low-energy excitations near the 
nodes, we choose local coordinate systems centered at $(\pm \pi/2, \pm \pi/2)$ 
adapted to each node, as shown in Fig.~\ref{coord}.

For any $A \neq 0$, it is evident that $E_{+,\k} \neq 0$ for any $\k$. The 
condition for $E_{-,\k} = 0$ can be reduced by some algebra to
\beq
\left( A^2 - \xi_{\k} \xi_{\k + \Q} - \Delta_{\k} \Delta_{\k + \Q} \right)^2 + \left( \xi_{\k} \Delta_{\k + \Q} - \xi_{\k + \Q} \Delta_{\k}  \right)^2 = 0
\label{nodeEq}
\eeq
Defining a local coordinate system $(k_1,k_2)$ about each of the collision 
points as shown in Fig.~\ref{coord}, to linear order in the momenta we find
\beq
\xi_{\k}  = v_F(k_0 + k_1), \; \Delta_{\k} = v_{\Delta} k_2, \; \xi_{\k + \Q} = - v_F(k_1 - k_0),  \;  \Delta_{\k + \Q} = -v_{\Delta} k_2
\eeq
where $v_F$ and $v_\Delta$ are the Fermi velocity and gap 
velocity, respectively, at any node (they are all identical due to the fourfold 
rotation symmetry). Substituting these in Eq.~(\ref{nodeEq}), we find 
that it reduces to
\beq
 \left( A^2 + v_F^2 k_1^2 + v_{\Delta}^2 k_2^2 - v_F^2 k_0^2 \right)^2 + (2 v_F 
v_{\Delta} k_o)^2 k_2^2 = 0.
\eeq
Thus, the nodes are located at
\beq
(k_1,k_2) = \left( \pm \sqrt{k_o^2 - A^2/v_F^2},0 \right),
\eeq
where the minus sign corresponds to the node within the reduced BZ, and the plus 
sign corresponds to the shadow node in the 2nd BZ. From this expression, it is 
obvious that no nodes exist for spiral order parameters $A$ beyond a critical 
value $A_c = v_F k_o$. One can visualize this as the node and the shadow note 
approaching each other and annihilating at $(k_1,k_2) = (0,0)$, resulting in a 
gap in the quasiparticle spectrum for $A \geq A_c$. Note that this nodal 
collision leads to a half-filled insulator.

\section{Derivation of thermal conductivity for co-existing N\'{e}el order and 
superconductivity in the clean limit}
For $\Q = (\pi,\pi)$ we note that $\vf(\k + \Q) \approx - \vf(\k)$, up to 
O$(k_0^2)$, and $\vd(\k + \Q) = - \vd(\k)$. The velocity matrix $\mathbf{V}_{\k} 
$ can hence be simplified to:
\beq
\mathbf{V}_{\k}  =  \begin{pmatrix}
\v_{1\k} & 0 \\
 0 &   \v_{2\k} \end{pmatrix}, \text{ where } \v_{1\k} =  \vf(\k) \tau_3 + \vd(\k) \tau_1, \text{ and }  \v_{2\k} =  -\vf(\k) \tau_3 + \vd(\k) \tau_1 \nn
\eeq
This allows to rewrite the trace in Eq.~(\ref{KappaLimcAF}) in terms of 
$G_{den}$, defined in Eq.~(\ref{ImGR}) (again, using labels 1 for $\k$ and 2 
for $\k + \Q$) at each nodal collision point,
\beq
 \text{Tr}\left[ G_R^{\prime \prime}(\k, 0)\mathbf{V}_{\k} G_R^{\prime 
\prime}(\k, 0)\mathbf{V}_{\k}\right] &=&  2 \Gamma^2 \bigg[ (A^2 +  \Gamma^2 + 
\xi_1^2 + \Delta_1^2)^2  + (A^2 +  \Gamma^2 + \xi_2^2 + \Delta_2^2)^2 \bigg] 
\left( \vf \vf + \vd \vd \right)/G_{den}^2 \nn 
 & & - 4 A^2 \Gamma^2  \bigg[ \left( (\xi_1 + \xi_2)^2 - (\Delta_1 + \Delta_2)^2 
\right)\bigg] \left( \vf \vf - \vd \vd \right)/G_{den}^2 \nn
 \label{trEqn}
\eeq

Defining a local coordinate system $(k_1,k_2)$ about each of the collision points as shown in Fig.~\ref{coord}, we linearize the different terms in the Hamiltonian:
\beq
\xi_1  = v_F(k_0 + k_1), \; \Delta_1 = v_{\Delta} k_2, \; \xi_2 = - v_F(k_1 - k_0),  \;  \Delta_2 = -v_{\Delta} k_2
\eeq
Note that the numerator in Eq.~(\ref{trEqn}) is proportional to $\Gamma^2$, 
which goes to zero in the clean limit. Therefore, the most important 
contributions come from the region of $k$ space where the denominator also goes 
to zero. In the linearized approximation described above, we find 
that:
\beq
G_{den} =  \left( A^2 + v_F^2 k_1^2 + v_{\Delta}^2 k_2^2 - v_F^2 k_0^2 \right)^2 + (2 v_F v_{\Delta} k_o)^2 k_2^2 + O(\Gamma_o^2)
\eeq
Therefore, one can see that $G_{den} = 0$ in the clean limit, only when $k_2 = 
0$ and $k_1^2 = k_o^2 - A^2/v_F^2$. Therefore, beyond a critical strength of the 
AF order amplitude, i.e, for $A > A_c = v_F k_o$, there is no solution. With 
regards to the spectrum, this corresponds to the scenario with gapped 
quasiparticles in the commensurate case, and therefore in the clean limit the 
conductivity is equal to zero at $T=0$. In the dirty limit when $\Gamma \gtrsim 
\Delta$, this is not the case as the disorder induced self-energy modifies 
quasiparticle spectral weights and closes the gap.   

As $\Gamma_0^2 \rightarrow 0$, the terms in O$(\Gamma_0^4)$ in the integral can 
be safely neglected, and the terms proportional to $\Gamma_0^2$ can be replaced 
by their values at the point where the denominator vanishes, i.e, $k_2 = 0$ and 
$k_1^2 = k_0^2 - A^2/v_F^2$. Within this approximation, we obtain
\beq
 \text{Tr}\left[ G_R^{\prime \prime}(\k, 0)\mathbf{V}_{\k} G_R^{\prime 
\prime}(\k, 0)\mathbf{V}_{\k}\right]  &=&  \Gamma^2  \frac{2 A_c^2 ( A_c^2 - 
A^2) \left( \vf \vf + \vd \vd \right) - A^2 A_c^2 \left( \vf \vf - \vd \vd 
\right)}{[\Gamma^2 A_c^2 + f ]^2}, \text{ where } \nn 
 f &=&  \frac{\left( A^2 + v_F^2 k_1^2 + v_{\Delta}^2 k_2^2 - A_c^2 \right)^2}{4} +  A_c^2 v_{\Delta}^2 k_2^2
\eeq
In the following we evaluate the diagonal conductivity and pick the 
$i^\text{th}$ component of the velocities, $i \in \{x,y\}$. Since in the 
coordinate system chosen, $\vf$ and $\vd$ are parallel to either $\hat{k}_1$ or 
$\hat{k}_2$, so we have:
\beq
2 \left( \vf \vf \pm \vd \vd \right)_{ii} = v_F^2 \pm v_{\Delta}^2 
\eeq
Rescaling the momenta by defining $\tilde{q}_1 = v_F k_1$ and $\tilde{q}_2 = 
v_{\Delta} k_2$, and multiplying by a factor of four for the four pairs of 
nodal points in the BZ (every point has the same 
contribution), we obtain
\beq
\frac{\kappa_{ii}(\Omega \rightarrow 0,T)}{T} & = & \frac{k_B^2 (v_F^2 + 
v_{\Delta}^2)}{3 v_F v_{\Delta}} \int \frac{d^2\tilde{q}}{2\pi} \frac{\Gamma^2 
A_c^2 ( A_c^2 - A^2/2)}{\left( \Gamma^2 A_c^2 + \left( A^2 + \tilde{q}^2 - 
A_c^2 \right)^2/4 +  A_c^2 \tilde{q}_2^2 \right)^2} \nn 
&  & - \frac{k_B^2 (v_F^2 - v_{\Delta}^2)}{3 v_F v_{\Delta}} \int 
\frac{d^2\tilde{q}}{2\pi} \frac{\Gamma^2 A_c^2 A^2/2}{\left( \Gamma^2 A_c^2 + 
\left( A^2 + \tilde{q}^2 - A_c^2 \right)^2/4 +  A_c^2 \tilde{q}_2^2 \right)^2}
\label{kappaInt}
\eeq
for the diagonal conductivity. The integrals in Eq.~(\ref{kappaInt}) can be 
analytically evaluated in the clean limit. We first cast the integrals in terms 
of dimensionless variables $\gamma = \Gamma_o/A_c$, $q_i = \tilde{q}_i/A_c$ and 
$\alpha = A/A_c$, measuring energy in units of $A_c = k_0 v_F$.
\beq
I_1 &=& \left( 1 - \frac{\alpha^2}{2} \right) \int \frac{d^2 q}{2\pi} \frac{\gamma^2 }{\left( \gamma^2 + \left( 1 + q^2 - \alpha^2 \right)^2/4 +  q_2^2 \right)^2}  \equiv  \left( 1 - \frac{\alpha^2}{2} \right) I_3 \nn
I_2 & = &  \frac{\alpha^2}{2} \int \frac{d^2 q}{2\pi} \frac{\gamma^2 }{\left( \gamma^2 + \left( 1 + q^2 - \alpha^2 \right)^2/4 +  q_2^2 \right)^2} \equiv \frac{\alpha^2}{2} I_3
\eeq
Now we change variables to $q_1 = 1 + x \cos \theta$, and $q_2 = x \text{ sin 
}\theta$. Then the denominator of the integral can be written as:
\beq
 \gamma^2 + \left( 1 + q^2 - \alpha^2 \right)^2/4 +  q_2^2 &=& \gamma^2 + x^2 + 
\left(\frac{x^2 + \alpha^2}{2}\right)^2 + x(x^2 + \alpha^2) \cos \theta \nn
 & = & \left( \tilde{\gamma}^2 + 1 + h^2 + 2 h  \cos \theta \right)x^2, \text{ 
where } \tilde{\gamma} = \gamma/x, h = \frac{x^2 + \alpha^2}{2x} \nn
\eeq
Plugging this back into the integral and shifting $\theta \rightarrow \theta + \pi$, we find:
\beq
I_3 &=&  \int_{0}^{\infty} \frac{dx \, x}{2 \pi }\int_{-\pi}^{\pi} d \theta 
\frac{\tilde{\gamma}^2 x^2}{x^4 \left( \tilde{\gamma}^2 + 1 + h^2 - 2 h  \cos 
\theta \right)^2} \nn
& = & \int_{0}^{\infty} \frac{dx \, x}{ \pi  } \int_{0}^{\pi} d \theta  
\frac{\tilde{\gamma}^2 x^2}{x^4 \left( \tilde{\gamma}^2 + 1 + h^2 - 2 h  \cos 
\theta \right)^2}  \nn
& = & \int_{0}^{\infty} \frac{dx }{ \pi  x } I_4
\eeq
where 
\beq
I_4  & \equiv & \int_{0}^{\pi} d \theta  \frac{\tilde{\gamma}^2}{\left( 
\tilde{\gamma}^2 + 1 + h^2 - 2 h  \cos \theta \right)^2} \nn
& = & \frac{2\pi (1 + h^2)}{(1 + h)^3} D(h-1, \tilde{\gamma}^2), \text{ with } D(u,v) \equiv \frac{v^2/2}{(u^2 + v^2)^{3/2}}
\eeq
where we have already used $\tilde{\gamma} \rightarrow 0$ to simplify the integral. Note that $D(u,v)$ vanishes in the limit of $v \rightarrow 0$ for all $u$ expect $u=0$, where it diverges. Moreover, it also satisfies:
\beq
\int_{-\infty}^{\infty} du D(u,v)  = 1
\eeq
Therefore, in the $\Gamma_o \rightarrow 0$ limit, which also corresponds to the second argument $\tilde{\gamma}^2 \rightarrow 0$, we can replace $D(h-1, \tilde{\gamma}^2)$ by $\delta(h-1)$. In this limit, we have:
\beq
I_4 = \frac{2\pi (1 + h^2)}{(1 + h)^3} D(h-1, \tilde{\gamma}^2) \; \overset{ \tilde{\gamma}^2 \rightarrow 0}{\longrightarrow} \; \frac{\pi}{2} \, \delta(h-1)
\eeq
Finally, we can plug back $I_4$ into $I_3$ and evaluate the sum over the delta-function:
\beq
I_3  & = & \int_{0}^{\infty} \frac{dx }{ \pi  x }\; \frac{\pi}{2}\, \delta\left( \frac{x^2 + \alpha^2}{2x} -1 \right) \nn
& = & \frac{1}{2} \int_{0}^{\infty} dx \left[ \frac{\delta(x - x_{+})}{\sqrt{1 - \alpha^2}} +  \frac{\delta(x - x_{-})}{\sqrt{1 - \alpha^2}} \right], \text{ where } x_{\pm} = 1 \pm \sqrt{1 - \alpha^2} \nn
& = & \frac{1}{\sqrt{1 - \alpha^2}} \; \text{ provided } \alpha < 1, \text{ and } 0 \text{ otherwise}
\eeq
Putting all of this together, we arrive at the finite expression of the static 
diagonal thermal conductivity in the clean limit for coexisting N\'{e}el order 
and superconductivity:
\beq
\frac{\kappa_{ii}(\Omega \rightarrow 0,T)}{T} =  \frac{k_B^2}{3 v_F v_{\Delta}} \bigg[ \sqrt{1 - \alpha^2} \, v_F^2 + \frac{1}{\sqrt{1 - \alpha^2}} v_{\Delta}^2 \bigg] \Theta(1 - \alpha), \text{ where } \alpha = \frac{A}{A_c}
\eeq

\section{Particle-particle bubble in spiral antiferromagnet or algebraic charge 
liquid}

\begin{figure}
\centering
\includegraphics[width=0.6\linewidth]{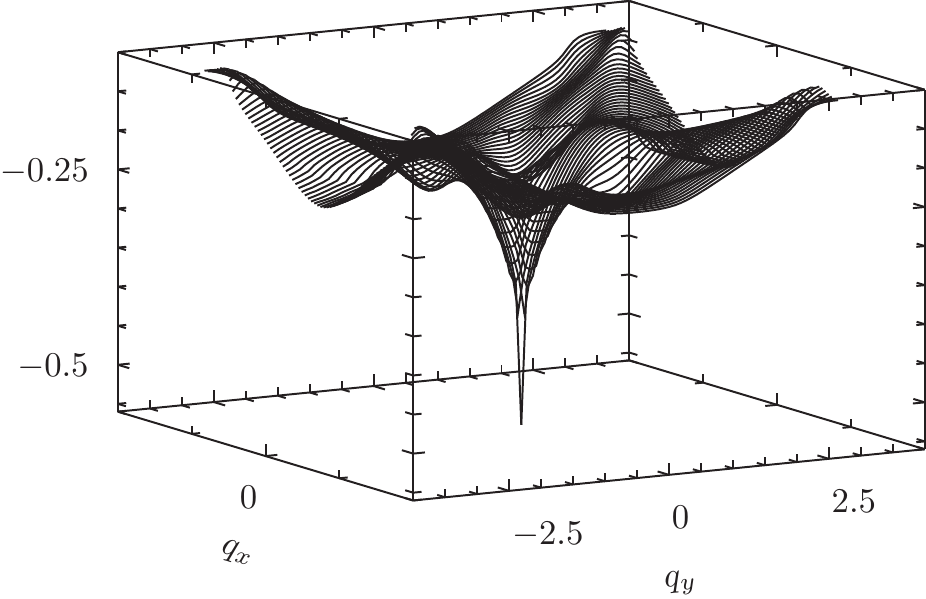}
\caption{Momentum dependence of the $d$-wave particle-particle bubble for $t' = 
 -0.35$, $A =   0.51$, $\eta = 0.08$ and $p = 0.09$.}
\label{fig:LPPd}
\end{figure}

In this paper, as well as in Refs.~\cite{Yamase2016, Sushkov2004}, it was 
assumed that Cooper pairs in a spiral antiferromagnetic state have vanishing 
total momentum. In order to substantiate this assumption, we compute the 
particle-particle bubble in a spiral state. Its momentum dependence at 
vanishing bosonic frequency is given by
\begin{align}
	L_\text{PP}(\bs q) =& - \int \frac{d^3 k}{(2\pi)^3} \Bigl\{f(\bs k + 
\tfrac{\bs q}{2})^2 G_{++}(k_0, \bs k) G_{--}(-k_0, \bs k + \bs Q) \\
	&+ f(\bs k + \tfrac{\bs 
q}{2}) f(\bs k + \bs Q + \tfrac{\bs q}{2}) G_{+-}(k_0, \bs k) 
G_{-+}(-k_0, \bs k + \bs Q)\Bigr\}. 
\nonumber
\end{align}
where $G_{ij}$ are the components of Eq.~\eqref{eq:G_iAF_metal}. This result can 
be rewritten as
\begin{align}
	=& - \int \frac{d^3 k}{(2\pi)^3} \Bigl\{f(\bs k + 
\tfrac{\bs q}{2})^2 G_{++}(k_0, \bs k) G_{++}(k_0, \bs k)^\ast 
\label{eq:LPP_d}\\
	&+ f(\bs k + \tfrac{\bs q}{2}) f(\bs k + \bs Q + \tfrac{\bs q}{2}) 
G_{+-}(k_0, \bs k) G_{+-}(k_0, \bs k)^\ast\Bigr\}
\nonumber
\end{align}
by exploiting the definition of the components of the propagator. Evaluation of 
the frequency integral yields
\begin{equation}
\begin{split}
=& -\int \frac{d^2 k}{(2\pi)^2} \sum_{i = \pm, j = \pm} \epsilon_i 
\epsilon_j \frac{1 - n_F(E_{\bs k + \bs q, i}) - n_F(E_{\bs k,j})}{(E_{\bs k + 
\bs q, i} + E_{\bs k, j})(E_{\bs k + \bs q, +} - E_{\bs k + \bs q, -})(E_{\bs k, 
+} - E_{\bs k, -})}\\
	& \times \Bigl\{f(\bs k + \tfrac{\bs q}{2})^2 (E_{\bs k + \bs q, i} - 
\xi_{\bs k + \bs q + \bs Q})(E_{\bs k, j} - \xi_{\bs k + \bs Q}) + f(\bs k + 
\tfrac{\bs q}{2}) f(\bs k + \bs Q +  \tfrac{\bs q}{2}) A^2\Bigr\}
\end{split}
\end{equation}
where $\epsilon_\pm = \pm 1$. In Fig.~\ref{fig:LPPd} we show numerical results 
for the particle-particle bubble in the $d$-wave channel. It is very strongly 
peaked at $\bs q = 0$, as assumed in Refs.~\onlinecite{Yamase2016, Sushkov2004}. This 
is already suggested by the functional form of $L_\text{PP}(\bs q)$ when 
written as in Eq.~\eqref{eq:LPP_d}. The momentum dependence does not possess a 
four-fold rotation symmetry, as expected in a spiral state.

\bibliographystyle{apsrev4-1_custom}
\bibliography{spiralAFtransport}

\end{document}